\documentclass[conference,compsoc]{IEEEtran}

\ifCLASSOPTIONcompsoc
  \usepackage[nocompress]{cite}
\else
  \usepackage{cite}
\fi

\usepackage{amsmath,amssymb,amsfonts,amsthm}
\usepackage{algorithmic}
\usepackage[ruled,vlined]{algorithm2e}
\usepackage{graphicx}
\usepackage{subfig}
\usepackage{xcolor}
\usepackage{caption}
 \usepackage{url}

\newtheoremstyle{mystyle}%                % Name
  {}%                                     % Space above
  {}%                                     % Space below
  {}%                                     % Body font
  {}%                                     % Indent amount
  {\bfseries\itshape}%                            % Theorem head font
  {.}%                                    % Punctuation after theorem head
  { }%                                    % Space after theorem head, ' ', or \newline
  {}%                                     % Theorem head spec (can be left empty, meaning `normal')
\theoremstyle{mystyle}
\newtheorem{theorem}{Theorem}[subsection]
\newtheorem{definition}[theorem]{Definition}
\newtheorem{deduction}[theorem]{Deduction}

\newtheorem{proposition}[theorem]{Proposition}

\newtheoremstyle{mystyle1}%                % Name
  {}%                                     % Space above
  {}%                                     % Space below
  {}%                                     % Body font
  {}%                                     % Indent amount
  {\itshape}%                            % Theorem head font
  {.}%                                    % Punctuation after theorem head
  { }%                                    % Space after theorem head, ' ', or \newline
  {}%                                     % Theorem head spec (can be left empty, meaning `normal')
\theoremstyle{mystyle1}
\newtheorem{apx-theorem}{Theorem}[subsection]
\newtheorem{apx-deduction}[apx-theorem]{Deduction}

\newenvironment{functionality}[1][htb]
  {
   \begin{algorithm}
  }{\end{algorithm}}

\begin{document}
\pagestyle{plain}

\title{ Threshold-Protected Searchable Sharing: \\ Privacy Preserving Aggregated-ANN Search for Collaborative RAG }

\author{\IEEEauthorblockN{Ruoyang Rykie Guo} \IEEEauthorblockA {gruoyang@stevens.edu}}

\maketitle

\begin{abstract}
LLM-powered search services have driven data integration as a significant trend. However, this trend's progress is fundamentally hindered, despite the fact that combining individual knowledge can significantly improve the relevance and quality of responses in specialized queries and make AI more professional at providing services. Two key bottlenecks are private data repositories' locality constraints and the need to maintain compatibility with mainstream search techniques, particularly Hierarchical Navigable Small World (HNSW) indexing for high-dimensional vector spaces.
In this work, we develop a secure and privacy-preserving aggregated approximate nearest neighbor search (SP-A$^2$NN) with HNSW compatibility under a threshold-based searchable sharing primitive. A sharable bitgraph structure is constructed and extended to support searches and dynamical insertions over shared data without compromising the underlying graph topology. 
The approach reduces the complexity of a search from $O(n^2)$ to $O(n)$ compared to naive (undirected) graph-sharing approach when organizing graphs in the identical HNSW manner.

On the theoretical front, we explore a novel security analytical framework that incorporates privacy analysis via reductions. The proposed leakage-guessing proof system is built upon an entirely different interactive game that is independent of existing coin-toss game design. Rather than being purely theoretical, this system is rooted in existing proof systems but goes beyond them to specifically address leakage concerns and standardize leakage analysis — one of the most critical security challenges with AI's rapid development.

\end{abstract}

\section{Introduction}

%\textcolor{blue}{[RAG limits/ challenge of the problem is -- why hard] }

As LLM-search systems scale to new heights, leveraging data integration from diverse individual and institutional sources empowers AI agents (e.g., GPTs) to decode complex professional contexts with enhanced accuracy and depth, especially within domain-specific fields such as biomedical laboratory research, independent research institutions, and expert-level query-response (QA) environments.
As private data is commonly stored in isolated and confidential local servers, significant privacy concerns prevent these specialized domains from sharing and integrating their critical data resources, limiting the realization of collaborative multi-user LLM-search platforms.  While existing single-user/multi-tenancy RAG \cite{DBLP:conf/nips/LewisPPPKGKLYR020} architectures help LLM chatbots access internal private data, they fail to support multi-user knowledge sharing platforms in a collaborative pattern, where data is remained confidential for each individual data owner without physically extracting the data from its original location.

To realize such a collaborative RAG environment requires an \textit{aggregated approximate nearest neighbors} (A$^2$NN) proximity search, given that standard RAG systems rely on ANN similarity search as their core mechanism for retrieving relevant contexts. While considering that existing ANN search techniques adopted in RAG systems heavily depend on \textit{hierarchical navigable small world} (HNSW) \cite{DBLP:journals/pami/MalkovY20} indexing, the mainstream approach for high-dimensional vector indexing, this dependency makes the Aggregated-ANN problem extremely challenging under security and privacy requirements.

In the realm of cryptographic protection techniques, the series of \textit{multi-party computation} (MPC) \cite{DBLP:journals/dam/Maurer06} techniques seemingly offers intuitively viable solutions that enable the aggregated setting by keeping local data in place while allowing participating users (i.e., parties) to jointly perform search calculations. However, the HNSW indexing method involves multilayer graphs, and this sophisticated structure requires multiple rounds of interactions if directly calculating each vertex of the graphs via MPC protocols. This introduces overwhelming complexity since each vertex corresponds to high-dimensional vector representations. In comparison, fully homomorphic encryption is unsuitable due to efficiency concerns. 
Another category of non-cryptographic approaches is inherently insufficient under this problem context, such as differential privacy  or anonymous technologies, since authoritative data in specialized domains demands the highest security standards and cannot be utilized in real-world applications without rigorously proven security guarantees. Existing secure search schemes based on the \textit{searchable encryption} (SSE) \cite{DBLP:conf/sp/SongWP00} technical line cannot simultaneously satisfy both graph-based indexing requirements \cite{DBLP:conf/sp/Servan-Schreiber22} and distributed private calculation demands in this setting. No methods are currently designed for either direct HNSW structures or indirect graph-based searches that can be adapted to HNSW indexing for vector search over encrypted data.

\textsc{Contributions.} 
In this work, we develop SP-A$^2$NN, a secure and privacy-preserving approximate nearest neighbors search that is compatible with HNSW indexing and supports distributed local data storage across individual users.
To achieve this goal, we first formally formulate the dynamic searchable sharing threshold primitive for the SP-A$^2$NN problem. A pattern combining arithmetic and secret sharing is utilized to perform distance comparisons during searches, while a sharable bit-graph structure is designed to minimize complexity, significantly decreasing search complexity from \textit{quadratic} to \textit{linear} compared to distributing original HNSW graphs in shared format. Interestingly, after transforming from an original undirected graph (i.e., with selective preservation of the original graph topology) to a bit-graph while using the proposed two rule designs \textit{at-hand-detour} and \textit{honeycomb-neighbor}, the searching walks remain nearly identical performance as in the original graphs.

On the theoretical analysis perspective, we adopt a novel quantifiable reduction-based security analytical framework that incorporates leakage analysis for validating SP-A$^2$NN's security and privacy. The critical need for introducing a new leakage-guessing proof system arises from a fundamental gap on leakage between existing proof systems based on adaptive security and their application to encrypted search schemes. Current approaches address this gap through leakage functions that capture non-quantifiable states and calculate detailed leakage ranges under threat models where attackers possess varying knowledge (typically derived from prior datasets, index, or their interconnections) to make leakage measurable. In this work, we claim that this leakage gap can be simulated in security environments through standardized calculations via formal reductions, while this still represents an uncharted theoretical field, the formal definitions and presentations are not entirely complete.
A high-level overview of such a proof system by reduction is illustrated in Appendix \ref{apx:framework}

%Below, we first outline the intuitions and structure of this leakage-guessing proof system.

\section{Background}

\subsection{ANN Search} \label{subsec:ann}
An appropriate nearest neighbor (ANN) search finds elements in a dataset that are approximately closest to a given query. 
The algorithm takes as input a query $\mathbf{q}$, a dataset of vectors $D$ and other parameters, and outputs approximate nearest-neighbor IDs.
We begin with revisiting the basic ANN search for processing a query in brute-force way in Functionality \ref{alg:ann search}. A threshold $\theta$ constrains the query range by setting either the maximum allowable neighbor distance or desired count of vectors to be returned. 
Next, we explore the HNSW indexing method that organizes high-dimensional vectors (e.g., embeddings) in a hierarchical structure to accelerate search.

\begin{functionality}[t]
\renewcommand{\thealgocf}{\ref{subsec:ann}.1}
\caption{ANN Search} 
\label{alg:ann search}
\begin{algorithmic}[1]  
\renewcommand{\algorithmicrequire}{\textbf{Input:}} 
      \REQUIRE 
    query $\mathbf{q}$, query threshold $\theta$ for either distance/range or nearest neighbor count, vector dataset $D=\{\mathbf{v}_i\}_{ 1\leq i \leq |D|}$, and other param\\-eters, i.e.,  \textit{distance computation metric $\mathsf{Distance}$, vector dimension $d$ with $\mathbf{q}, \mathbf{v}_i \in \mathbb{R}^d$.}
\renewcommand{\algorithmicrequire}{\textbf{Output:}} 
      \REQUIRE 
        Nearest neighbors. 
 \renewcommand{\algorithmicrequire}{\textbf{Brute-Force Procedure:}} 
      \REQUIRE 
       \STATE $\mathbf{a} \leftarrow$ nearest neighbor to $\mathbf{q}$ in  $DB$ via brute-force search
     \IF{$\mathsf{Distance}(\mathbf{a}, \mathbf{q}) > \theta$} 
       \STATE \qquad the client outputs $Null$ and $\perp$.
       \ELSE
       \STATE \qquad the client outputs a vector $\mathbf{a}$ and $\perp$.
       \ENDIF
    \end{algorithmic}
\end{functionality}

\begin{functionality}[t]
\renewcommand{\thealgocf}{\ref{subsec:hnsw_for_ann}.1}
\caption{HNSW Indexing Sketch for ANN Search} \label{alg:hnsw_for_ann}
    \begin{algorithmic}[1]
\renewcommand{\algorithmicrequire}{\textbf{Input}:} 
      \REQUIRE 
      query $\mathbf{q}$, query threshold $\theta$ for either distance/range or nearest neighbor count, vector dataset $D =\{\mathbf{v}_i\}_{1\leq i\leq |D|}$ with index $\mathcal{I}$,
      and other parameters, i.e.,  \textit{distance metric $\mathsf{Distance}$, vector dimensions $d$ with $\mathbf{q}, \mathbf{v}_i \in \mathbb{R}^d$}.
\renewcommand{\algorithmicrequire}{\textbf{Output}:} 
      \REQUIRE 
      Nearest neighbors.
\renewcommand{\algorithmicrequire}{\textbf{HNSW Procedure}:} 
      \REQUIRE
       \STATE \ $\mathsf{Path} \leftarrow$ a routing path that connects layers of $\mathcal{I}$ by probabilistically skipping vectors according to their distances (near or far)
       \STATE $\mathbf{a} \leftarrow$ nearest neighbor to $\mathbf{q}$ in the $0$th later found via $\mathsf{Path}$
       \STATE \ \ \textbf{if} $\{\mathsf{Distance}(\mathbf{a},\mathbf{q}) > \theta \}$ \ \textbf{then} 
       \STATE  \qquad the client outputs $Null$ and $\perp$.
        \STATE \  \ \textbf{else} 
       \STATE  \qquad the client outputs a vector $\mathbf{a}$ and $\perp$.
    \end{algorithmic}
\end{functionality}

\subsection{HNSW Graph-Based Indexing for ANN Search }\label{subsec:hnsw_for_ann}
A hierarchical navigable small world (HNSW) algorithm
organizes a dataset $D$ using a multilayered graph-based index $\mathcal{I}$, with each layer being an undirected graph with vertices as data elements. The layers compose a hierarchical structure from top to down where each upper layer is extracted from the layer below it with a certain probability. Within each layer, graph construction follows a distance-priority way in which elements closer in distance are more likely to be connected through an edge. Generating such a multi-layer index is a dynamic process of inserting dataset elements (i.e., vectors) one by one from the top layer down to the bottom layer, where the bottom layer contains the complete dataset.

When processing a query, the algorithm traverses from top layer to the most bottom layer until a query range of appropriate nearest neighbors is reached.
Given a query element, HNSW search finds the nearest element in each top layer (excluding the bottom layer). The nearest element found in the $L$th layer becomes the starting anchor for the next lower layer ($(L\text{-}1)$th),  then the search finds its nearest element in that layer; and this process continues layer by layer until reaching the bottom layer, where the final nearest neighbors are identified within a specific range. Finally, the IDs of neighbors satisfying the threshold $\theta$ are returned as the result. The search sketch is shown in Functionality \ref{alg:hnsw_for_ann}. We recommend referring to Fig. 1 of ref \cite{DBLP:journals/pami/MalkovY20} for a better understanding of the search process.
 In brief, the proximity graph structure replaces the probabilistic skip list \cite{Skip_list.wiki}, maintaining a constant limit on edges (i.e., connections) per layer, which enables HNSW search to achieve fast logarithmic complexity for nearest neighbor queries, even with high-dimensional vector data.

Following real-world database architecture that
organizes data via index, it is established that a database consists of two parts: dataset and index as
\begin{equation}
\begin{split}
    DB & = D + \mathcal{I}.
\end{split}
\end{equation}

\section{Redefine Problem}
In this section, we define the system, security and threat model of a secure and privacy-preserving aggregated approximate nearest neighbor (SP-A$^2$NN) search problem.

\subsection{System, Security and Threats Model of SP-A$^2$NN  Search Scheme} \label{subsec:system, security, threats model}

\textit{Participating Parties.} A party can be a client such as a service provider (e.g., biomedical laboratory) using collaborative computing services, or an endpoint user (e.g., platform-agnostic worker) seeking to establish a collaboration network with others. 
Take the RAG frameworks for example, each party configures a database typically as vectors based on their individual knowledge (e.g., files) to leverage external AI retrieval services such as language models.
The computing task is to retrieve relevant knowledge across all parties, creating a collaborative knowledge database to let language models easily draw upon when producing answers. For brevity, our framework focuses only on outputting the retrieved data, excluding the process of parties forwarding the results to a language model.

A set $\mathcal{U}$ of $n$ parties participate in executing an aggregated SP-A$^2$NN search. Imagine that all parties integrate their data and jointly establish/update a global index, maintaining an idealized collaborative database
$C\text{-}DB$ together, in which a global index $C\text{-}\mathcal{I}$ organizes a unified dataset $C\text{-}D$ across parties. Searches using this global index are completed through interactions among parties, with each element of the unified dataset accessible via a unique pointer in the global index regardless of this element ownership. The final search result aggregates the queried nearest neighbors from all parties. The structure is represented as 
\begin{equation}
\begin{split}
    C\text{-}DB = C\text{-}D + C\text{-}\mathcal{I}
\end{split}
\end{equation}

\textit{Security and Threat Model.} In a SP-A$^2$NN search, participating parties do not trust one another and seek to keep their individual databases confidential from other parties. We consider \textit{honest-but-curious} security environments, where parties follow the protocol honestly but may attempt to infer other parties' data during execution. While this can be extended to prevent \textit{active} adversaries through additional processes consistency verification, we omit this from the current work. While the threat model traditionally concerns attackers' prior knowledge, in this work, we employ a privacy triplet setting to connect standardizable threat patterns with leakage analysis. The objective is to make privacy analysis as the foundation for the security analysis framework.

\section{Preliminaries}
\newcommand{\q}{\quad}

In this section, we define a basic cryptographic primitive, dynamic searchable sharing threshold (SST), for formulating the problem of SP-A$^2$NN search. Under this primitive, we provide related security definitions and related constructions in Sec \ref{subsec:dynamic SST}, along with the existing cryptographic building blocks used in this realization in Sec \ref{subsec:building blocks}. Sec \ref{subsec:privacy triplet} defines privacy triplet.

\subsection{Dynamic SST} \label{subsec:dynamic SST}
Dynamic SST evolves from dynamic SSE capabilities for searches and updates (such as insertion and deletion), adapting its definitions to work in a database environment where data is distributed across multiple separate parties.

\textit{Conceptual Settings.}
As in SSE schemes, $EDB$ denotes the encrypted database that combines encrypted index and encrypted data blocks (i.e., storage units), but with expanded scope in dynamic SST. We introduce $C$-$EDB$ as an abstract construct, representing an idealized collaborative database in encrypted form that integrates an unified encrypted dataset (i.e., $C\text{-}ED$) with a global encrypted index (i.e., $C\text{-}E\mathcal{I}$), that is
\begin{equation}
\label{equa:C_EDB}
   C\text{-}EDB = C\text{-}ED+C\text{-}E\mathcal{I}.
\end{equation}
\noindent
From a real perspective, $C$-$EDB$ integrates separate dataset segments, along with index segments,  distributed among and maintained by parties,
\begin{align} 
C\text{-}EDB &  = \sum_{1}^{n} EDB_i \ \\ & =
\sum_{1}^{n} ED_i + \sum_{1}^{n} E\mathcal{I}_i
\end{align}
\noindent
where $EDB_i$ is the portion of $C$-$EDB$ that is physically stored in party $u_i$, including data and index segment, $ED_i$ and $E\mathcal{I}_i$ respectively.

\textbf{Dynamic SST Definition.} A dynamic SST problem $\Sigma =\{\mathsf{Setup}, \mathsf{Search}, \mathsf{Update}\}$ is comprised of interactive protocols as:

$\mathsf{Setup}(1^\lambda)\rightarrow K, \sigma, C\text{-}EDB$: It takes as input database $DB$ and $\lambda$, the computational security parameter of the scheme (i.e., security should hold against attackers running in time $\approx 2^\lambda$). The outputs are collectively maintained by all parties. 
$K$ is secret key of the scheme,
analogous to the arithmetic protection applied to data (e.g., the constructed polynomial formula in \textit{Shamir's} secret sharing \cite{DBLP:journals/cacm/Shamir79}). $\sigma$ is an chronological state agreed across parties, and $C\text{-}EDB$ is an encrypted  (initially empty) collaborative database.

$\mathsf{Search}(K, \sigma, \mathbf{q}; C\text{-}EDB) \rightarrow C\text{-}EDB(\mathbf{q})$: It represents a protocol for querying the collaborative database. We assume that a search query $\mathbf{q}$ is initiated by party $v$. The protocol outputs $C\text{-}DB(\mathbf{q})$, 
meaning that the elements that are relevant to $\mathbf{q}$ (i.e., appropriate nearest neighbors in vector format) are returned.

$\mathsf{Insert/Delete}(K, \sigma, in; C\text{-}EDB)\rightarrow K, \sigma, C\text{-}EDB$:  It is a protocol for inserting an element $in$ into (or deleting it from) the collaborative database.
The element $in$ is a vector owned by the party who requests an update. The protocol ends with a new state where all parties jointly confirm if $C\text{-}EDB$ contains the element $in$ or not.

The above definitions extend the APIs of common dynamic SSE \cite{DBLP:conf/ccs/BostMO17} to adapt the database structure, specifically representing data storage blocks (e.g., dataset $C\text{-}ED$) and index (e.g., $C\text{-}E\mathcal{I}$). The $\mathsf{Search}$ algorithm's result shows which nearest neighbors are retrieved in response to a given query, while the process is independent of how parties subsequently forward the results to a language model (Sec \ref{subsec:system, security, threats model}).

\textsc{Quantifiable Correctness.} A dynamic SST problem $\Sigma = (\mathsf{Setup}, \mathsf{Search}, \mathsf{Update})$ is correct if it returns the correct results for any query with allowable deviation. 

Correctness is a relative term that quantifies identical search results by comparing them to a baseline search as reference, where this baseline is generally a search over plaintext data under the same index. While searching in any applied secure scheme, result deviation inevitably occurs, making it difficult to maintain the same identical search results that an Enc/Dec oracle can achieve. Therefore, we introduce a concept of deviation to define correctness below.

\textsc{Quantifiable Security.} 
A dynamic SST problem $\Sigma = (\mathsf{Setup}, \mathsf{Search}, \mathsf{Update})$ is secure with bounded leakage if it is proven to satisfy an allowable privacy budget of a certain value that can be standardized for measurement.

Both the privacy and threshold-based security analysis of SST are captured simultaneously using a reduction-based leakage-guessing proof system: The \textsc{Ideal} experiment expresses the layer where the basic security scheme achieves provable threshold-based security, while the \textsc{Real} experiment represents any applied scheme (such as our proposed SP$^2$ANN scheme) that, as the outermost layer, must capture leakage. Although direct reduction \textit{w.r.t} security from \textsc{Real} to \textsc{Ideal} can identify threshold-based security, it cannot locate where/what level of leakage occurs. 
A new \textsc{Mirror} environment is then introduced as an intermediary bridge that enables comparison with the \textsc{Ideal} environment for threshold-based security analysis and with the \textsc{Real} environment for leakage analysis.

\begin{definition}[\textit{$\Delta$-bounded Deviation-Controlled Correctness of Dynamic SST}]  
\label{def:sst_cor}
A dynamic SST problem $\Pi$ is $\Delta$-correct 
\textit{iff} for all efficient $\mathcal{A}$, there exists a stateful efficient $S$, such that
\begin{equation} 
\begin{split}
  & \quad \mathbf{Adv}^\text{SST-Cor}_{\Pi, \mathcal{A}} (\lambda, \rho) = \\
  & \mathbf{Adv}^\text{SST-Cor}_{\Pi_\text{bas}, \mathcal{A}} (\lambda) +  \mathbf{Adv}^\text{SST-Cor}_{\Pi_M, \mathcal{A}} (\lambda) +  \mathbf{Adv}^\text{SST-Cor}_{\Pi, \mathcal{A}} (\lambda, \rho)
\end{split}
\end{equation}
where the first two functions satisfy

\begin{equation}
\label{equa:cor_Pi_m}
\begin{split}
   \mathbf{Adv}^\text{SST-Cor}_{\Pi_M, \mathcal{A}} (\lambda)  = & \{\textsc{Mirror}_{\Pi_M}^\mathcal{A}(\lambda)\} \quad \equiv   \\
   &  \{\textsc{Ideal}_{\Pi_\text{Bas}}^\mathcal
   A(\lambda)\} = \mathbf{Adv}^\text{SST-Cor}_{\Pi_\text{bas}, \mathcal{A}} (\lambda)
\end{split}
\end{equation}  
and are \textit{negligible}, and
\begin{equation} 
  \mathbf{Adv}^\text{SST-Cor}_{\Pi, \mathcal{A}} (\lambda, \rho) = \{\textsc{Real}_{{\Pi}}^\mathcal{A} (\lambda)\} - 
   \{\textsc{Sim}_{\Delta(\Pi, \Pi_M),S}^\mathcal
   A(\lambda, \rho)\} 
\end{equation}
is a \textit{unnegligible} function in terms of the allowable deviation $\rho$. $\{\cdot\}$ means the probability that adversary wins in the experiment.
\end{definition}

\begin{definition}[\textit{$\mathcal{L}(\epsilon)$-bounded Threshold Security of Dynamic SST}]  
\label{def:sst_sec}
A dynamic SST problem $\Pi$ is $\mathcal{L}$-secure 
\textit{iff} for all efficient $\mathcal{A}$, there exists a stateful efficient $S''$, $S'$ and $S$, such that
\begin{equation} 
\begin{split}
& \quad \mathbf{Adv}^\text{SST-Sec}_{\Pi, \mathcal{A}} (\lambda, \epsilon) =  \\
 & \mathbf{Adv}^\text{SST-Threshold}_{\Pi_\text{Bas}, \mathcal{A}} (\lambda)  + \mathbf{Adv}^\text{SST-Threshold}_{\Pi_M, \mathcal{A}} (\lambda) +   \mathbf{Adv}^\text{SST-Privacy}_{\Pi, \mathcal{A}} (\lambda, \epsilon) 
\end{split}  
\end{equation}
\noindent
where
\begin{equation}
\label{equa:adv_bas}
   \mathbf{Adv}^\text{SST-Threshold}_{\Pi_\text{Bas}, \mathcal{A}, S''} (\lambda)  = \{\textsc{Ideal}_{\Pi_\text{Bas}}^\mathcal{A}(\lambda)\} - \{\textsc{Sim}_{\mathcal{L}(\Pi_\text{Bas}, SS), S''}^\mathcal{A}(\lambda)\} 
\end{equation}  
\begin{equation}
\label{equa:adv_mirror}
   \mathbf{Adv}^\text{SST-Threshold}_{\Pi_{M}, \mathcal{A}, S'} (\lambda)  = \{\textsc{Mirror}_{\Pi_{M}}^\mathcal{A}(\lambda)\} - \{\textsc{Sim}_{\mathcal{L}(\Pi_{M}, \Pi_\text{Bas}), S'}^\mathcal{A}(\lambda)\} 
\end{equation}  
are both \textit{negligible} functions, and
\begin{equation} 
\label{equa:adv_privacy_real}
  \mathbf{Adv}^\text{SST-Privacy}_{\Pi, \mathcal{A}, S} (\lambda, \epsilon) = \{\textsc{Real}_{{\Pi}}^\mathcal{A} (\lambda)\} - 
   \{\textsc{Sim}_{\mathcal{L}(\Pi, \Pi_M),S}^\mathcal
   A(\lambda, \epsilon)\} 
\end{equation}
is a \textit{unnegligible} function in terms of the allowable privacy budget $\epsilon$. $\{\cdot\}$  follows the same meaning.
\end{definition}

\textsc{Experiments Definition.} Importantly, the above experiments  extend existing query-response games for adaptive data security. For instance, $\{\textsc{Real}_{{\Pi}}^\mathcal{A} (\lambda)\}$ (Def \ref{def:sst_sec}.(\ref{equa:adv_privacy_real})) defines adversary's adaptive security advantage against encrypted data in a real scheme $\Pi$, with this advantage is verified through game-based experiments. In this work,  we assume that part of the security has been validated; therefore, we omit the query-response game experiments for adaptive security while looking forward a little bit. Instead, we particularly focus on a gap in current security proof systems: privacy simulation.

In Definition \ref{def:sst_sec}, 
$\{\textsc{Sim}_{\mathcal{L}(\Pi, \Pi_M),S}^\mathcal
A(\lambda, \epsilon)\} $ is $\mathcal{A}$'s advantage on an environment for simulating $\Pi$. With the same logic, this advantage is established based on but extends beyond an adaptive security environment, where a simulator $S$ simulates a reduction from $\Pi$ to $\Pi_M$, and the output of this environment is the leakage $\mathcal{L}(\Pi, \Pi_M)=\mathcal{L}(\epsilon)$. Analogously,
$ \{\textsc{Sim}_{\mathcal{L}(\Pi_{M}, \Pi_\text{Bas}), S'}^\mathcal{A}(\lambda)\}$ is an environment for simulating $\Pi_M$ that is provided by $\{\textsc{Mirror}^\mathcal{A}_{\Pi_M}\}$; and $\mathcal{A}$'s advantage is calculated via $S'$'s simulation of the reduction from $\Pi_M$ to $\Pi_\text{Bas}$, where the leakage $\mathcal{L}(\Pi_M, \Pi_\text{Bas}) = \mathcal{L}(\lambda)$ meaning that it is allowable.

In Definition \ref{def:sst_cor}, $ \{\textsc{Sim}_{\Delta(\Pi, \Pi_M),S}^\mathcal
A(\lambda, \rho)\} $ is an environment for simulating $\Pi$ that is provided by $\{\textsc{Real}_{{\Pi}}^\mathcal{A} (\lambda)\}$ \textit{w.r.t} correctness, 
where $\Delta$ is a function for capturing result deviation between $\Pi $ and $ \Pi_M)$. 
Of particular note, 
$\{\textsc{Mirror}_{\Pi_{M}}^\mathcal{A}(\lambda)\}$ establishes the correctness baseline, meaning it satisfies complete correctness equivalence with $\{\textsc{Ideal}_{\Pi_\text{Bas}}^\mathcal
A(\lambda)\}$.

\textbf{Basic Construction.} Let a $(t,n)$-threshold secret sharing configuration $SS$ serve as an encryption scheme of $(\mathsf{Enc}, \mathsf{Dec})$, $F_1$ be polynomial formula (i.e., keys) for producing shares and $F_2$ be an arithmetic circuit for calculating ciphers. We have our basic static construction $\Pi_{SS}$ for the dynamic SST problem in Fig \ref{fig:basic_scheme}.

\begin{theorem}
\label{theo:Pi_ss to ss}
A basic scheme $\Pi_{SS}$ is correct and threshold-secure \textit{iff} the $(t,n)$-threshold secret sharing mechanism $SS$ is information-theoretically secure. 
\end{theorem}

\textbf{Mirror Construction.} Let
a $(t,n)$-threshold secret sharing configuration $SS$ serve as an encryption scheme of $(\mathsf{Enc}, \mathsf{Dec})$, and $\mathcal{I}\text{-}hnsw$ be the HNSW index to organize $C\text{-}EDB$. $F_1$ and $F_2$ use the same representations in $\Pi_{SS}$.
We have our mirror construction $\Pi_{SS}^{\mathcal{I}\text{-}hnsw}$ in Fig \ref{fig:mirror_scheme}.

\begin{theorem}
\label{theo:Pi_ss_hnsw to Pi_ss}
A mirror scheme $\Pi^{\mathcal{I}\text{-}hnsw}_{SS}$ is correct and $\mathcal{L}$-secure \textit{iff} $\Pi_{SS}$ is threshold-secure and the reduction from $\Pi_{SS}^{\mathcal{I}\text{-}hnsw}$ to $\Pi_{SS}$ \textit{w.r.t leakage} is $\mathcal{L}(\Pi_{SS}^{\mathcal{I}\text{-}hnsw},\Pi_{SS})$-secure.
\end{theorem}

The proofs for Theorem \ref{theo:Pi_ss to ss} and Theorem \ref{theo:Pi_ss_hnsw to Pi_ss} are provided in Appendix \ref{apx_proof:Pi_bas to ss} and  \ref{apx_proof:Pi_m to Pi_bas} respectively.

\begin{figure*}[t]
\renewcommand{\baselinestretch}{1.1}
\small
\centering
  \fboxsep=2pt
  \fbox{ 
  \begin{minipage}[t]{0.32\dimexpr\linewidth-2\fboxsep-40\fboxrule}
  \begin{algorithmic}[1]
 \renewcommand{\algorithmicrequire}{\underline{$\mathsf{Setup}(1^\lambda, \sigma)$}}
 \REQUIRE 
    \STATE $sd_i \overset{\$}{\leftarrow}  \{0,1\}^\lambda$ allocate list $L$
    \STATE Initiate Counter $\sigma: c \leftarrow 0$
    \STATE $K_i \leftarrow  F_1(sd_1, c)$ 
    \STATE Add $K_i$ into list $L$ (in lex order)
    \STATE Output $K=(K_i,  \sigma)$
    \end{algorithmic}
  \end{minipage}
  
\begin{minipage}[t]{0.38\dimexpr\linewidth-2\fboxsep-40\fboxrule}
   \begin{algorithmic}[1]
 \renewcommand{\algorithmicrequire}{\underline{$\mathsf{Insert}(K, \sigma, \mathbf{q}; C\text{-}EDB)$}}
 \REQUIRE 
    \STATE (\textit{party} $u$) $\{\mathbf{q}_u\}_\mathcal{U} \leftarrow \mathsf{Enc}(\mathbf{q}, K_1)$ 
    \vspace{8pt}
     \STATE Set \par
     $C\text{-}E\mathcal{I} \leftarrow C\text{-}E\mathcal{I}.\mathsf{Add}(\{loc(\mathbf{q}_u)\}_\mathcal{U})$; \par $C\text{-}ED \leftarrow C\text{-}ED.\mathsf{Add}(\{ \mathbf{q}_u\}_\mathcal{U})$; \par
      $\sigma: c$++
     \STATE Output \par $C\text{-}EDB= (C\text{-}E\mathcal{I}, C\text{-}ED, \sigma)$
     \end{algorithmic}
  \end{minipage} 
  
  \begin{minipage}[t]{0.36\dimexpr\linewidth-2\fboxsep-40\fboxrule}
  \begin{algorithmic}[1]
 \renewcommand{\algorithmicrequire}{\underline{$\mathsf{Search}(K, \sigma, \mathbf{q}; C\text{-}EDB)$}}
 \REQUIRE 
    \STATE (\textit{party} $v$) $\{\mathbf{q}_u\}_\mathcal{U} \leftarrow \mathsf{Enc}(\mathbf{q}, K_2)$ 
      \vspace{8pt}
    \STATE On input $\{\mathbf{q}\}_\mathcal{U}$ and  \par
   $C\text{-}EDB = (C\text{-}E\mathcal{I}, C\text{-}ED, \sigma)$
    \STATE For $c=0$ until $\mathsf{BruteForce}$ return $\perp$, \par
    $\{\mathbf{v}_u\}_\mathcal{U}  \leftarrow $ \par \quad \q  $
 \mathsf{BruteForce}(C\text{-}E\mathcal{I}; C\text{-}ED, \{\mathbf{q}\}_\mathcal{U}, F_2)$
   \STATE $\mathbf{v} \leftarrow \mathsf{Dec}(\{\mathbf{v}\}_\mathcal{U}, K)$
   \STATE Output $\mathbf{v}$
   \end{algorithmic}
  \end{minipage}}
\caption{Basic Scheme $\Pi_{SS}$}
\label{fig:basic_scheme}
\end{figure*}

\begin{figure*}[t]
\renewcommand{\baselinestretch}{1.1}
\small
\centering
  \fboxsep=2pt
  \fbox{
  \begin{minipage}[t]{0.32 \dimexpr\linewidth-2\fboxsep-40\fboxrule}
  \begin{algorithmic}[1]
 \renewcommand{\algorithmicrequire}{\underline{$\mathsf{Setup}(1^\lambda, \sigma)$}}
 \REQUIRE 
    \STATE $sd_i \overset{\$}{\leftarrow}  \{0,1\}^\lambda$ allocate list $L$
    \STATE Initiate Counter $\sigma: c \leftarrow 0$
    \STATE $K_i \leftarrow  F_1(sd_1, c)$ \par
    \STATE Add $K_i$ into list $L$ (in lex order)
    \STATE Output $K=(K_i,  \sigma)$
  \end{algorithmic}
  \end{minipage}

\begin{minipage}[t]{0.38 \dimexpr\linewidth-2\fboxsep-40\fboxrule}
  \begin{algorithmic}[1]
 \renewcommand{\algorithmicrequire}{\underline{$\mathsf{Insert}(K, \sigma, \mathbf{q}; C\text{-}EDB)$}}
 \REQUIRE 
    \STATE (\textit{party} $u$) $\{\mathbf{q}_u\}_\mathcal{U} \leftarrow \mathsf{Enc}(\mathbf{q}, K_1)$ 
    \vspace{8pt}
     \STATE Set \par
     $C\text{-}E\mathcal{I} \leftarrow $  \par \quad \q $C\text{-}E\mathcal{I}.\mathsf{Add}(\mathcal{I}\text{-}{hnsw}:\{loc(\mathbf{q}_u)\}_\mathcal{U})$; \par $C\text{-}ED \leftarrow C\text{-}ED.\mathsf{Add}(\{ \mathbf{q}_u\}_\mathcal{U})$; \par
      $\sigma: c$++
     \STATE Output \par $C\text{-}EDB= (C\text{-}E\mathcal{I}, C\text{-}ED, \sigma)$
  \end{algorithmic}
  \end{minipage} 

  \begin{minipage}[t]{0.36 \dimexpr\linewidth-2\fboxsep-40\fboxrule}
  \begin{algorithmic}[1]
 \renewcommand{\algorithmicrequire}{\underline{$\mathsf{Search}(K, \sigma, \mathbf{q}; C\text{-}EDB)$}}
 \REQUIRE 
    \STATE (\textit{party} $v$) $\{\mathbf{q}_u\}_\mathcal{U} \leftarrow \mathsf{Enc}(\mathbf{q}, K_2)$ 
      \vspace{8pt}
    \STATE On input \par
    $\{\mathbf{q}\} \leftarrow v: \mathsf{Enc}(\mathbf{q})$
    \\$C\text{-}EDB = (C\text{-}E\mathcal{I}, C\text{-}ED, \sigma)$
    \STATE For $c=0$ until $\mathsf{HNSW}$ return $\perp$, \par
    $\{\mathbf{v}_u\}_\mathcal{U}  \leftarrow $ \par \quad \q  $
 \mathsf{HNSW}(C\text{-}E\mathcal{I}; C\text{-}ED, \{\mathbf{q}\}_\mathcal{U}, F_2)$
   \STATE $\mathbf{v} \leftarrow \mathsf{Dec}(\{\mathbf{v}\}_\mathcal{U}, K)$
   \STATE Output $\mathbf{v}$
  \end{algorithmic}
  \end{minipage} }
  
  \caption{Mirror Scheme $\Pi_{SS}^{\mathcal{I}\text{-}hnsw}$}
    \label{fig:mirror_scheme}
 \vspace{-3pt}
\end{figure*}

\subsection{Cryptographic Building Blocks} \label{subsec:building blocks}

\textit{Shamir's $t$-out-of-$n$ Secret Sharing Scheme.} Within this mechanism, any subset of $t$ shares enables recovery of the complete secret $s$ that has been divided into $n$ parts, while any collection of up to $t-1$ shares yields no information about $s$. The generation of shares is parameterized over a finite field $\mathbb{F}$ of size $l>2^k$ (i.e., $k$ is the security parameter of the scheme), where, e.g., $\mathbb{F}=\mathbb{Z}_p$ for some public prime $p$ \footnote{The selection of a prime $p$ is constrained by some public integer $r$ with $l=p^r$, $p^r > 2^k$}. In our scheme, parties share their data as vectors. As a result, we constrain this field size parameter by $l \geq \mathsf{Max}(2^k, \left \lceil 10^\rho \right \rceil, n)$, where $10^\rho$ represents the scaling factor applied to a vector.
The scheme is composed of two algorithms $\mathbf{\mathsf{SS}}=(\mathsf{SS.Share}_n^t, \mathsf{SS.Recon}_n^t)$, one for sharing a secret with parties and the other for reconstructing a share from a subset of parties.

The sharing algorithm $\mathsf{SS.Share}_n^t(s) \rightarrow \{(u, s_u)\}_{u\in \mathcal{U}}$  takes as input a secret $s$, a set $\mathcal{U}$ with size $|\mathcal{U}|=n$ for parties, a threshold $t \leq n$, and it produces a set of shares, representing a party $u$ holds a share $s_u$ for all parties in $\mathcal{U}$. The reconstruction algorithm $ \mathsf{SS.Recon}_n^t(\{(u,s_u)\}_{u\in \mathcal{V}}) \rightarrow s$ inputs a threshold $t$ and shares related to a set $\mathcal{V} \subseteq  \mathcal{U}$ with  $|\mathcal{V}|\geq t$, and outputs the secret $s$ as a field element.
    
\textsc{Correctness.} A $t$-out-of-$n$ secret sharing scheme $\mathsf{SS}$ correctly shares a secret $s$ if it always reconstructs $s$. 

\textsc{Security.} A $t$-out-of-$n$ secret sharing scheme privately shares a secret $s$ if $\forall s,s'\in \mathbb{F}$ and any $\mathcal{V} \subseteq \mathcal{U}$ with $|\mathcal{V}|<t$, there exists the view of parties in $\mathcal{V}$ during an execution of $\mathsf{SS}$ for sharing $s$ and that view for $s'$, such that 
\begin{equation}
    \{\textsc{View}^{\mathsf{SS}}_{u\in \mathcal{V}}(\{(u, s'_u)_{u\in \mathcal{U}}\})\} \equiv \{\textsc{View}^{\mathsf{SS}}_{u\in \mathcal{V}}(\{(u, s_u)_{u\in \mathcal{U}}\})\},
\end{equation}
\noindent
where $\{(u, s'_u)_{u\in \mathcal{U}}\} \leftarrow \mathsf{SS.Share}_n^t(s')$, $\{(u, s_u)_{u\in \mathcal{U}}\} \leftarrow \mathsf{SS.Share}_n^t(s)$, and $\equiv$ denotes computational indistinguishability \textit{(bounded by distributions).}

\subsection{Privacy Triplet for Leakage Analysis}\label{subsec:privacy triplet}

An intuition is that the ratio of inferrable data (e.g., based on prior knowledge) to the complete database provides a measure of leakage severity. We define prior knowledge as information already known to an adversary excluding publicly available information such as open-source indexing algorithms. When an adversary attempts to deduce additional information from a private database, we assume their prior knowledge is limited to a single, randomly chosen data entry. This approach allows us to measure the fraction of the database that becomes exposed when following deduction paths from a single, randomly selected data entry. The resulting ratio of inferrable data provides a standardized metric for comparing leakage across different security schemes. 

For formulating this, we redefine privacy leakage via an analytical framework, named \textit{Privacy Triplet}. This triplet defines three interfaces (I-III) of measurable leakage with dependently progressive strength as follows.  A complete inference trajectory, traversing from the starting interface (I) to the final interface (III), traces a linking path to its impacted inferrable data, beginning from a single, randomly selected data entry.
A complete trajectory following a privacy triplet is defined as:

\textbf{I. \textit{Data-to-Index} privacy interface $\mathcal{L^I}$.}
It leaks the index nodes that match a chosen data item.

\textbf{II. \textit{Index-to-Index} privacy interface $\mathcal{L^I}$.} It leaks the index nodes that can be deduced through other nodes already linked to the chosen data item.

\textbf{III. \textit{Index-to-Data} privacy interface $\mathcal{L}^D$.} It leaks additional data (or indirect information of data) that can be connected to the inferred index nodes from I and II.

\begin{definition}[\textit{Privacy Triplet}]
\label{def:privacy triplet}
Given any search scheme (e.g., dynamic SST $\Sigma$) build on an encrypted database $EDB$, a privacy triplet standardizes the leakage disclosure of $EDB$ by taking an individual, randomly chosen data item $w$ through a complete I-III trajectory as:
\begin{align*}
  \quad  & \textbf{I-\textit{Data-to-Index}}: \mathcal{L}^\mathcal{I}(w)= \mathcal{L}'(\textbf{DB-Inx}(w), w).\\
  \quad  & \textbf{II-\textit{Index-to-Index}}: \mathcal{L}^\mathcal{I}(w')= \mathcal{L}'(\textbf{DB-Inx}(w'), \textbf{DB-Inx}(w)).\\
  \quad  & \textbf{III-\textit{Index-to-Data}}: \mathcal{L}^D(w)= \mathcal{L}'(w', \textbf{DB-Inx}(w)) 
\end{align*}
\noindent
where $\mathcal{L}'$ is a stateless function.
\end{definition}

\textit{Cooperating with Existing Privacy Norms.}
An privacy triplet creates an analytical approach for bounding leakage severity, making it not conflict with established metrics that identify \textit{patterns} of leak-inducing behaviors.  
Existing leakage patterns describe leak-inducing behaviors occurring during search, update and access operations. Update and access patterns have been relatively well explored and defined. \textit{Access patterns} captures the observable sequence of data locations that are accessed during searches. \textit{Update patterns} are typically associated with forward and backward privacy, which address the leakage incurred by earlier and later updates (i.e., insertions and deletions). Existing works define \textit{search patterns} in a more flexible way to explore how correlations between previously executed and subsequent queries affect information exposure.
We illustrate possible locations within the leakage-guessing analytical framework where these patterns can be integrated, as shown in the Appendix \ref{apx:framework}. 
\section{Technical Intuitions} \label{sec:technical_intuitions}

We lay the technical foundation and preparations in this section for constructing an efficient aggregated approximate nearest neighbor search scheme (SP\text{-}A$^2$NN) with security and privacy guarantees, detailed in Sec \ref{sec:sp-a2nn}. 
This involves a novel storage structure termed \textit{bitgraph} along with its essential functionalities in Sec \ref{subsec:bitgraph} to enable subsequent effective aggregated searching. In Sec \ref{subsec:necessity_bitgraph}, we examine the efficiency dilemma that emerges when naively distributing HNSW graphs (undirected graphs) to construct a sharable index, thereby justifying our bitgraph approach. Through complexity analysis comparing unmodified graphs with bitgraphs for executing aggregated queries, we show that bitgraphs achieve a reduction from \textit{quadratic} to \textit{linear} complexity.

By design, the bitgraph structure significantly decreases the number of invocations of Shamir's secret sharing by using minimal information to convey the complete structure of HNSW graphs. The additional optimizations further implement search/update functionalities by introducing the most minimal changes possible based on the bitgraph framework.

\subsection{Establishing the Critical Need for Bitgraph} \label{subsec:necessity_bitgraph}
In what follows, we first analyze theoretical arguments on computational complexity that unavoidably arises when sharing undirected graphs without proper conversions, and then discuss the inherent tensions between this cost, search functionality, and efficiency.

\textsc{Complexity Analysis}.
We begin with a scenario wherein an undirected graph, comprising vertices and their connecting edges, is to be distributed across multiple parties in such a way that all participants (at least $t$) possess sufficient information to reconstruct the complete graph.
A graph's topological structure is captured in its vertex connection pattern, with each edge serving as a link between two vertices. Thus, when quantifying the complexity of graph sharing, the metric is the minimum number of vertices that must be distributed in shares and exchanged among parties to convey the graph's complete structure.

If we consider the sharing operation on a vertex as a one-time pad encryption, then this vertex cannot be traced back after it has been checked during searching. This means connections related to a vertex must be collaboratively recorded when sharing it. In consequence, the link between two vertices must be shared repeatedly among parties even only considering minimum sharing patterns. We define such structures as \textit{sharable sets} for representing the complete structure of a graph, and we only focus on the minimum sharable set of any given graph.

\begin{definition}[\textit{Minimum Sharable Set (MSS)}] \label{def:MSS}
    For sharing any undirected graph $G=(V,E):$ $\{V$ as a set  of vertices  and $E$ for a set of edges$\}$ among parties, there exists a minimum sharable set $S$ that contains all distinct directed connections between vertices, and we have the size of this set satisfies twice the number of edges, that is  
    $\mathsf{Size}(S)= 2|E|$.
\end{definition}

Based on the definition on MSS, the complexity analysis for sharing an undirected graph becomes more straightforward. We validate this through several logical deductions.

\begin{deduction}
For any undirected graph $G=\{V,E\}$ and its minimum sharable set $S$,
when sharing $G$ among $n$ parties using a $t$-out-of-$n$ secret sharing mechanism $\mathsf{SS.Share}_n^t$, the computational complexity is:
   \begin{equation}
       O(\mathsf{SS.Share}_n^t(G))= \mathsf{Size}(S) \cdot n \cdot O(\mathsf{SS.Share}_n^t(\cdot)),
   \end{equation}
\end{deduction}
\noindent
where $O(\mathsf{SS.Share}_n^t(\cdot))$ denotes the computational cost of an invoking the sharing algorithm for each element in $S$. The relation holds because this complexity is proportional to the count of invoking sharing algorithms, which is proportional to size of its minimum sharable set. (Note that when we discuss complexity in this work, we are specifically measuring only the computational burden placed on a single party.)

\begin{figure}[h]
\centering
\includegraphics[scale = 0.57]{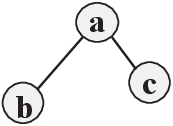}
\caption{A simple graph example}
\label{fig:exg_simple_graph}
\end{figure}

Take the graph with two edges in Fig \ref{fig:exg_simple_graph} as an instance, its minimum sharable set is  $\{\mathbf{a}\text{-}\mathbf{b}, \mathbf{a}\text{-}\mathbf{c}, \mathbf{b}\text{-}\mathbf{a}, \mathbf{c}\text{-}\mathbf{a}\}$. In this case, the computational cost of conveying the complete structure of it among parties is linear to at least four times the count of sharing operations.

Whereas, when characterizing the computational complexity of sharing a complete graph, it is more appropriate to express this in terms of the number of vertices rather than the number of edges. This metric is suitable because the vertex count directly corresponds to both the count of sharing operations and the database size (where each vertex represents a vector record). By analyzing the relations between edges and vertices  (Def \ref{def:MSS}), we establish an upper bound on the size of this minimum set, that is $\mathsf{Size}(S)\leq |V|^2$. (Referring to a basic deduction\footnote{For any  graph $G=(V,E)$, the maximum degree of any vertex is at most $|V|-1$, and then the total number of edges $|E|$ cannot exceed $\frac{|V|(|V|-1)}{2}$. } in the context of graph theory \cite{Graph_theory.wiki}.)

\begin{deduction}
For any undirected graph $G=\{V,E\}$ and its minimum sharable set $S$, when sharing $G$ among $n$ parties using $\mathsf{SS.Share}_n^t$, assuming any vertex $\mathbf{v}$  in the vertex space satisfies the format $\mathbf{v}\in \mathbb{R}^d$, the computational complexity  is:
\begin{equation}
   O(\mathsf{SS.Share}_n^t(G)) =  d|V|^2 \cdot n \cdot O(p^t),
\end{equation}
\end{deduction}
\noindent
where $d$ denotes vector dimension, $p$ is modular of the finite field arithmetic in which the mechanism $\mathsf{SS.Share}_n^t$ operates and $t$ is the degree of the polynomial used for calculating shares.

\textit{Reconciling Search Functionality Challenge.} 
Efficiency is impeded by the \textit{quadratic} invocations for generating shares, and although not the primary factor, this contributes to significant computational flows during search operations.
The process of repeatedly sharing vertices and their association relations leads to \textit{quadratic} distance comparison calculations over shares, an essential requirement for search functionality to determine if a specific vertex is being queried.

Real-world searches over feature vectors commonly utilize three distance operators for similarity evaluation: \textit{Euclidean Distance}, \textit{Inner Product} and \textit{Cosine Similarity}.
Assuming the compatibility with existing arithmetic protocols to implement distance calculations over shares, selecting an arithmetic circuit that supports a full set of calculation operators to compute shares becomes an unavoidable step. However, the computational burden imposed by any arithmetic protocol that offering universal calculation capabilities (i.e., both addition and multiplication) is substantial enough to impact any single search, regardless of the specific protocol chosen — even without considering the additional authentication costs required to address malicious security threats.

We visualize the complexity of searching on shares through the following deduction.

\begin{deduction}
Let $c$ denote the number of vertices in a searching walk over any undirected graph $G =\{V,E\}$ with $c \leq |V|$. Then, when sharing $G$ among $n$ parties using $\mathsf{SS.Share}_n^t$ and based on previous deductions, the computational complexity of a search operation is:
 \begin{equation}
        O(\mathsf{Search\_on\_Shares}(\{u, G_u\}_{u \in \mathcal{U}})) = dc^2 \cdot n \cdot O(p^t) \cdot O(\mathsf{AC}),
 \end{equation}
\noindent
where $O(\mathsf{AC})$ represents the complexity of arithmetic circuits that are required for performing distance similarity and comparisons between two vertices.
\end{deduction}

In this work, we address the search efficiency dilemma under this problem by minimizing the number of arithmetic circuit invocations, rather than reducing the cost of each invocation. In the next subsection, we propose a novel storage structure, bitgraph, to support \textit{linear} complexity for sharing, calculating distance over shares, and reconstructing search results, making practically effective index and search systems possible.

\subsection{Sharable Bitgraph Structure} \label{subsec:bitgraph}

A sharable bitgraph is the key storage structure required for constructing a sharable index to realize aggregated ANN search. This bitgraph structure, derived from HNSW graphs (i.e., undirected graphs), maintains the integrity of original inserting and searching walks. 
The proposed bitgraph eliminates the process of sharing entire graphs with their full set of vertices and edges, significantly reducing complexity (i.e., sharing times) from \textit{quadratic} to \textit{linear}. 

\textit{Intuitions.} We explain the efficiency of bitgraph sharing by showing how it eliminates redundant vertex connection records through a strategic decomposition of graph information. This design significantly reduces complexity while preserving complete graph representation. A bitgraph replaces the traditional vertex-edge structure with four components: \textit{vertices}, \textit{sequences}, \textit{post-positive degrees}, and \textit{parallel branches}. The sequence component extracts the graph's fundamental structure: an ordered
path that visits all vertices where any vertex in this path keeps a single forward connection with its pre-positive vertex, while the path obviously and potentially omitting some edges. Post-positive degrees record only the backward connections of each vertex along this sequence. Conceptually, if we view a graph as a system of connected branches (i.e., subgraphs), the first three components fully describe individual branches without capturing inter-branch connections. The fourth component, parallel branches, records these branch-to-branch relationships. With all four components, we can completely reconstruct the original graph starting from any vertex.

\textit{Search Intuitions.} We introduce a conceptual intuition for searching: the search behaves like a walk that winds through the graph structure in a hexagonal honeycomb pattern, advancing toward deeper regions of a graph. This design leverages the previously described information to establish a natural bidirectional search trajectory, with one direction following forward connections and the other following backward connections, both guided by the established vertex sequences.

%which aligns with one-time pad protection by making 

\textit{Bitgraph Roadmap.} 
In what follows, we first discuss two pre-requisite definitions, subgraphs and its partition in undirected graphs and the isomorphism relations how graphs relate to bitgraphs. A bitgraph construction is then presented with a helpful example. 
Next, we provide algorithms for insertion and search operations on bitgraphs, with searches built on HNSW principles and enhanced with bitgraph-specific optimizations. Throughout, examples demonstrating these operations are provided. We conclude by establishing correctness proofs for search result consistency and analyzing the complexity of bitgraph sharing involved in constructing/searching bitgraphs across multiple shares.
(Note that graphs in this context have practical interpretations: a vertex represents a high-dimensional vector, while an edge corresponds to distance metrics between these vectors.)

\begin{definition} [\textit{Subgraph}] \label{def:sugraph}
    Any undirected graph with ordered vertices and edges  $G=\{V, E\}$ can be expressed as a set of subgraphs generated by applying a partitioning function $\Gamma$ as
    \begin{equation}
     G \overset{\Gamma}{=} \bigcup \mathrm{Subgraph}(G) \overset{\Gamma}{=} \underset{i}{\bigcup}(G_i),
    \end{equation}
\noindent
such that this union set contains the complete vertices (possibly repeated), edges, and the original ordered structure of $G$.  
\end{definition}

\begin{definition}[\textit{Bitgraph Isomorphism}] \label{def:bitgraph_isomorphism}
    A bitgraph isomorphism $f$ from an undirected graph $G$ to a bitgraph $H$ is a bijection (i.e., one-to-one correspondence) between the subgraph set of $G$ and the branch set of $H$, that is
    \begin{equation}
        f: \bigcup \mathrm{Subgraph}(G) \rightarrow \bigcup \mathrm{Branch}(H),
    \end{equation}
\noindent
such that each branch of $H$ is the image of exactly one subgraph of $G$. To further explore the isomorphism $f_i$, which maps the vertex and edge structure of a subgraph $G_i$ to its corresponding branch $H_i$, we introduce 
    \begin{equation}
        f_i: G_i \rightarrow H_i.
    \end{equation}
\noindent
Specifically, we define the branch set for a bitgraph $H$ as
\begin{equation}
     \underset{i}{\bigcup}(H_i) = \bigcup \mathrm{Branch}(H) =H.
\end{equation}
\end{definition}

\textbf{Bitgraph Construction.}
Given the above isomorphism between a bitgraph $H$ and undirected graph $G=\{V, E\}$,  $\{f_i: G_i \rightarrow H_i| G\overset{\Gamma}{=}\underset{i}{\bigcup}(G_i), H=\underset{i}{\bigcup}(H_i) \}$, we construct such a bitgraph by constructing a partition rule $\Gamma$ on its isomorphic graph $G$ and components of each branch $H_i$. 

The partitioning function $\Gamma$ operates on $G$ as follows: When traversing vertices according to their order, the function evaluates whether an edge exists between a vertex $\mathbf{v}_i$ and its next vertex $\mathbf{v}_{i+1}$ (i.e., adjacent in order). If no edge connects $\mathbf{v}_i$ and $\mathbf{v}_{i+1}$, then the vertex preceding $\mathbf{v}_{i+1}$ (denoted as $\mathbf{v}_j$) serves as a split vertex that generates a subgraph. This procedure is recursively applied to each resulting subgraph (replacing the original graph $G$ with the subgraph in the rule) until no vertices violate the connectivity rule. Note that within each subgraph, the partitioning rule remains consistent, but the vertex ordering refers to the sequence within the subgraph rather than in the original graph.

The composition of each branch is an ordered set $H_i = (V_i, seq(V_i), post\_d(V_i,V), par\_b(V_i))$ consisting of:
 
  \begin{itemize}
      \item   $V_i$, a set of vertices that forms a subset of $V$;
      \item $seq(V_i)$, a set containing the sequence of vertices in $H_i$;    
      \item $post\_d(V_i, V)$, a set containing post-positive degree of each vertex in $V_i$ with respect to the traversal sequence in $V$, defined as 
        \begin{equation}
            post\_d (V_i,V)=\{post\_d(y)| y=x,x\in V_i, y \in V\}.
        \end{equation}
    \item $par\_b(V_i)$, the set of branches in which a vertex from
 $V_i$ serves as the split vertex that creates a new branch.
  \end{itemize}

\begin{figure}[b]
\includegraphics[scale=0.44]{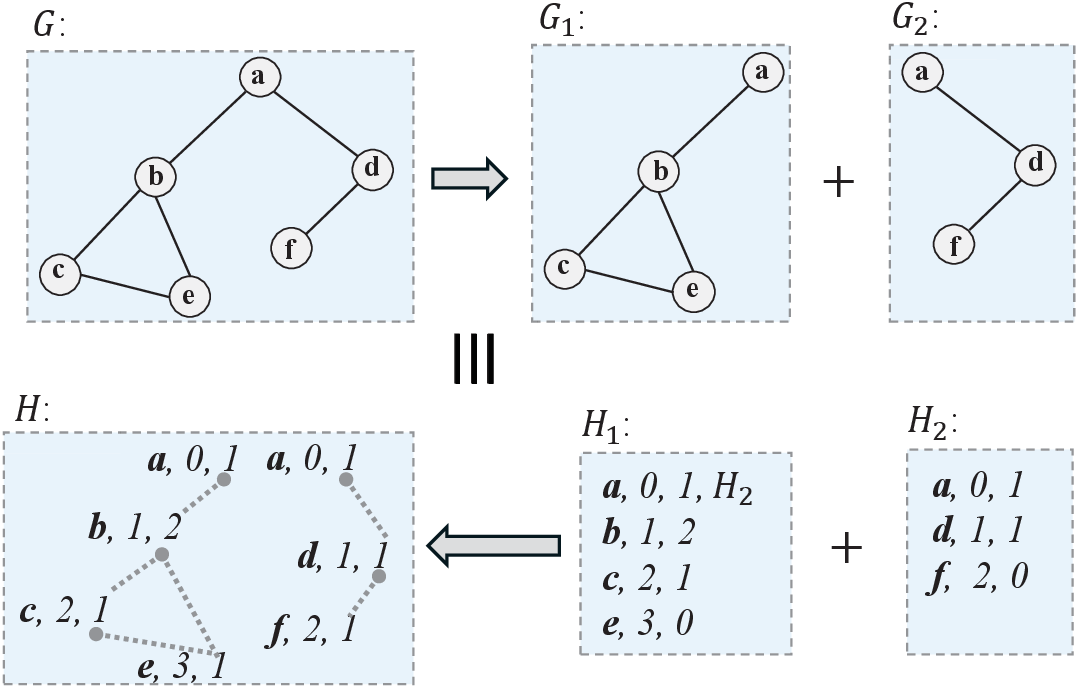}
\caption{A bigraph construction example}
\label{fig:exg_bitgraph_construct}
\end{figure}

\textit{Demonstrative Example.} To illustrate the construction, we present an example demonstrating how an undirected graph $G$ is partitioned and how its corresponding bitgraph $H$ is calculated. Consider an undirected graph $G$ with six vertices as shown in Fig \ref{fig:exg_bitgraph_construct}, where the alphabetic order (i.e., $\mathbf{a}, \mathbf{b}, \mathbf{c}, \mathbf{d}$, ...) represents the vertex ordering in $G$.
According to the partitioning function $\Gamma$, it can be observed that vertex $\mathbf{c}$ and its next vertex $\mathbf{d}$ are not adjacent in $G$'s vertex ordering. Therefore, the vertex preceding $\mathbf{d}$, namely vertex $\mathbf{a}$, serves as a split vertex. This creates a subgraph $G_2$ that follows the edge connection from $\mathbf{a}$ to $\mathbf{d}$.  

Let's examine this from a panoramic  time-sequential perspective. At this moment, vertices $\mathbf{a}, \mathbf{b}, \mathbf{c}$ comprise subgraph $G_1$, while vertices $\mathbf{a}, \mathbf{d}$ form subgraph $G_2$. 
When partition $\Gamma$ is applied recursively and independently to the subgraphs, we consider the established ordering within each subgraph, where $G_1$ contains vertices $\mathbf{a}, \mathbf{b}, \mathbf{c}$ in positions $0^{th}, 1^{st}, 2^{nd}$ respectively, while $H_2$ contains vertices $\mathbf{a}, \mathbf{d}$  in positions $0^{th}$ and $1^{st}$ respectively. Within $G_1$, since vertex $\mathbf{e}$ is the next vertex in sequence adjacent to $\mathbf{c}$ and they share a connecting edge, the incorporation of $\mathbf{e}$ into the subgraph extends $G_1$ rather than creating a new subgraph.  Likewise, the addition of vertex $f$ to $G_2$ extends $G_2$ without creating any new subgraph.
Consequently, $G$ is divided into exactly two subgraphs.

In calculating the branches of $H$ from the two subgraphs of
 $G$,  we follow the principle that $G_1$ yields $H_1$ and $G_2$ yields $H_2$ (as defined in Def. \ref{def:bitgraph_isomorphism}). To construct branches of $H$, such as branch $H_1$, we record triplet information from its isomorphic subgraph $G_1$.
 For instance, vertex $\mathbf{a}$, being the $0^{th}$ enter vertex in $H_1$ with one incident edge (connecting to its postpositive vertices) and creating a new branch $H_2$, is recorded as entry $(\mathbf{a}, 0, 1, H_2)$ where $\mathbf{a}$ already indicates its position in graph $G$'s ordered sequence; For vertex $\mathbf{b}$, which is the $1^{st}$ vertex in $H_1$ and connects to two post-positive vertices (i.e., $\mathbf{c}, \mathbf{e}$) without  producing a new branch, the recorded entry is  $(\mathbf{b}, 1, 2, \O)$. 
In the same manner, other vertices are converted to their triplets in $H_1$ and $H_2$. In particular, when vertex $\mathbf{g}$ forms connections with the tail vertexes (i.e., $\mathbf{e}, \mathbf{f}$) in both subgraphs $G_1$ and $G_2$ simultaneously, it is recorded in both $H_1$ and $H_2$. 
Examination of Fig \ref{fig:exg_bitgraph_construct} also reveals that when all branches in  $H$ are combined, they uniquely determine the reconstruction of the isomorphic graph $G$.

\textbf{Functionalities on Bitgraph Construction.} 
In the following text, we present algorithms for core bitgraph operations: $\mathsf{Bitgraph.Insert}$
and $\mathsf{Bitgraph.Search}$. These algorithms demonstrate how partition rules enable bitgraph representation of vertex insertion into a graph and how nearest neighbors are searched. We provide high-level intuitions for each algorithm design and visualize the algorithms using extreme cases where vertices are added/searched across bitgraph branches  for easier understanding.

The insertion algorithm's goal is to place a new vertex $\mathbf{q}$ in branch $H_i$ with pre-positive vertex set $W_i$, as shown in $\mathsf{Bitgraph.Insert}$ (Alg \ref{alg:bitgraph.insert}). This process operationalizes partition rule $\Gamma$ to decide vertex position in bitgraph, which depends on $W_i$ characteristics and whether a continuously adjacent vertex subset exists when traversing backwards. If found, vertices in this subset won't initiate new branches, but remaining vertices outside the subset will each create new branches. In the absence of such subsets, each vertex generates a new branch.  The input $W_i$ is commonly derived from the neighbor search results for vertex $\mathbf{q}$ before its insertion into $H_i$; we omit this process.

The search algorithm follows the basic logic of the original search in HNSW to replicate the searching walks for finding nearest neighbors of a vertex. However, parallel branches with cross-entering vertices make the original searching walks extremely difficult to maintain consistency when searching over bitgraph.  We begin with the HNSW search intuition and then show how to preserve the searching logic with minimal modifications, tracing the same searching walks to hold result consistency.

During traversal, HNSW search maintains two dynamic queues while traversing vertices: $C$, sequentially storing distinct vertices checked during search walks, and $W$, containing identified nearest neighbors. The path of search walks is determined by evaluating neighboring vertices to navigate deeper into the graph.
All distance comparisons utilize these queues, with $C$ consistently providing the vertex currently nearest to query vertex $\mathbf{q}$, and $W$ contributing the furthest vertex within query-range threshold $\theta$;  for example, vertices $\mathbf{c}$ and $\mathbf{f}$ respectively. Each comparison evaluates whether a new nearest neighbor exists by comparing distances $(\mathbf{c}, \mathbf{q})$ and $(\mathbf{f}, \mathbf{q})$ to update $W$. The search process terminates when, after finding sufficient nearest neighbors in $W$, the nearest vertex in $C$ is further than the furthest vertex in $W$.

\begin{figure}[b]
\includegraphics[width=2.9in, height =2.4in]{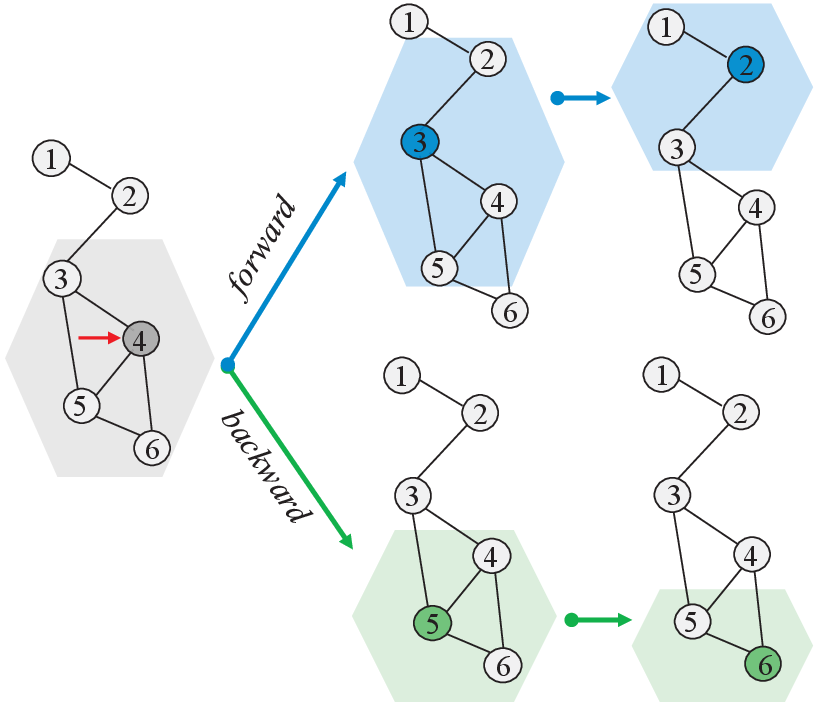}
\caption{A search trajectory through honeycomb neighbors on a branch}
\label{fig:exg_bitgraph_honeycomb_neighbors}
\end{figure}

To eliminate the uncertainty in search walk progression caused by cross-entering vertices, we modifies two places, and the complete algorithm is in $\mathsf{Bitgraph.Search}$ (Alg \ref{alg:bitgraph.search}): The first is the $\mathsf{Bitgraph.HoneycombNeighbors}$  algorithm (Alg \ref{alg:bitgraph.honeycomb_neighbors}), rewriting the vertex neighbor identification to governs which vertex enters the $C$ queue next. Through this algorithm, we identify all vertices \textit{honeycomb-adjacent} to the currently examined vertex $\mathbf{c}$, regardless of whether they reside in the current branch or in parallel branches (the latter scenario occurring when $\mathbf{c}$ functions as a split vertex). This method alone proves insufficient when search progress reaches a branch endpoint without triggering the termination condition, with searching still active.  To address this limitation, our second modification introduces an \textit{at-hand-detour} function (colored \textcolor{blue}{blue}, Lines 10-14) that reverts to the previously nearest vertex in queue $C$ when the examined vertex $\mathbf{c}$ is determined to be at its branch tail. This traceback approach establishes a specific pathway to vertices that should maintain connectivity in the original graph structure but have been segmented across different branches. In specific, we explain how Algorithm \ref{alg:bitgraph.honeycomb_neighbors} identifies \textit{honeycomb-adjacent} neighbors of its input vertex $\mathbf{c}$. On branch $H_i$, the honeycomb around vertex $\mathbf{c}$ consists of post-positive vertices (recorded in $post\_d$) and a single pre-positive vertex (recorded in $seq$). 
Besides its role within the current branch, we must consider cases where vertex $\mathbf{c}$ functions as a split vertex connecting to other forked branches (recorded in $par\_b$). In these cases, the honeycomb  involves only a single post-positive vertex that follows the current split vertex in sequence. Here, one step to the next vertex is enough to keep the search moving forward. 
A search trajectory example through honeycomb neighbors on a branch is visualized in Figure \ref{fig:exg_bitgraph_honeycomb_neighbors}, where the numerical values are the sequence ordering of vertices within the branch.

\begin{algorithm}[h]
\renewcommand{\thealgocf}{\ref{subsec:bitgraph}.2}
\caption{$\mathsf{Bitgraph.HoneycombNeighbors}(\mathbf{c}, H_i)$} 
\label{alg:bitgraph.honeycomb_neighbors}
 \begin{algorithmic}[1]
\renewcommand{\algorithmicrequire}{\textbf{Input}:} 
      \REQUIRE a vertex  $\mathbf{c}$,  its located branch $H_i$
\renewcommand{\algorithmicrequire}{\textbf{Output}:} 
      \REQUIRE neighbors of vertex $\mathbf{c}$
\renewcommand{\algorithmicrequire}{\textbf{Finding Neighbors across Branches} ---} 
      \REQUIRE
      \STATE $Neighbors \leftarrow \O$ // set of neighbors of vertex $\mathbf{c}$
    \STATE $\mathbf{c}\_par\_b \leftarrow $ get the parallel 
    branches set of $\mathbf{c}$ in $H_i$
    \STATE // if $\mathbf{c}$ is head vertex in $H_i$,\\ \quad \ $seq$ in line 5 starts from $\mathbf{c}.seq+1$ to $\mathbf{c}.post\_d$
     \STATE // if $\mathbf{c}$ is tail vertex in $H_i$,\\ \quad \ $seq$ in line 5 is assigned $\mathbf{c}.seq-1$
    \FOR{$seq\leftarrow  \mathbf{c}.seq-1, \mathbf{c}.seq+1 \ ...\ \mathbf{c}.seq + \mathbf{c}.post\_d $ in $H_i$} 
    \STATE $(\mathbf{v}, loc\_H_i) \leftarrow$ get the $seq^{th}$ vertex in $H_i$
    \STATE $Neighbors =  Neighbors \bigcup (\mathbf{v}, loc\_H_i)$
    \ENDFOR
    
    \FOR{each branch $H_j$ in $\mathbf{c}\_par\_b$}
   \STATE // get the next vertex of head vertex in $H_j$
    \STATE $(\mathbf{v}, loc\_H_j) \leftarrow$ get the $1^{st}$ vertex of $H_j$
    \STATE $Neighbors =  Neighbors \bigcup (\mathbf{v}, loc\_H_j)$
    \ENDFOR
    \RETURN $Neighbors$
\end{algorithmic} 
\end{algorithm}

\begin{algorithm}[h]
\renewcommand{\thealgocf}{\ref{subsec:bitgraph}.1}
\caption{$\mathsf{Bitgraph.Insert}(\mathbf{q}; H, (W_i, loc\_H_i))$} 
\label{alg:bitgraph.insert}
 \begin{algorithmic}[1]
\renewcommand{\algorithmicrequire}{\textbf{Input}:} 
      \REQUIRE a new vertex $\mathbf{q}$; a bitgraph $H$, and  $\mathbf{q}$'s pre-positive vertices set $W_i$ with an ordered sequence, its branch location $H_i$ (i.e., $W_i \subseteq H_i$).
\renewcommand{\algorithmicrequire}{\textbf{Output}:} 
      \REQUIRE
       update related entries of bitgraph $H$ after inserting $\mathbf{q}$ (i.e., add connections from neighbors $(W_i, loc\_H_i)$ to $\mathbf{q}$ in a given bitgraph.)
\renewcommand{\algorithmicrequire}{\textbf{Insert Procedure on $H_i$} ---} 
      \REQUIRE 
      \STATE $S \leftarrow \O$ // set of new split vertices
      \STATE $\{H_j: H_j \leftarrow \O \}$ // set of new branches produced from the split vertices in $H_i$
     % \IF{ $H_i = \O$} 
     % \STATE $(\mathrm{Entry}) \ \mathbf{q}\_seq,\mathbf{q}\_post\_d, \mathbf{q}\_par\_b \leftarrow 0, 0,\O$
     % \STATE $H_i \leftarrow H_i \bigcup \mathrm{thisEntry}.\mathbf{q}$
     % \ENDIF
 \STATE $T \leftarrow$  get set of vertices having continuous sequences from $W_i$ in a  back-to-front order
      \STATE $\mathbf{t} \leftarrow$ get last element from  $T$
       \STATE $\mathbf{h} \leftarrow$ get last element from  $H_i$
        \IF{$ \mathbf{t} = \mathbf{h} $}    
            \FOR{ each vertex $\mathbf{v} \in T$}
            \STATE $\mathrm{thisEntry}.\mathbf{v}\_post\_d \leftarrow \mathrm{thisEntry}.\mathbf{v}\_post\_d+1$ // update $\mathbf{v}$'s entry in $H_i$  
            \ENDFOR
       \STATE $(\mathrm{Entry}) \
 \mathbf{q}\_seq,\mathbf{q}\_post\_d, \mathbf{q}\_par\_b \leftarrow |H_i|, 0,\O$
        \STATE $H_i \leftarrow  H_i\bigcup \mathrm{thisEntry}.\mathbf{q}$  // add $\mathbf{q}$ to the tail of $H_i$
             \IF{$W_i/T \neq \O$}
          \STATE $S \leftarrow S\bigcup W_i/T $
        \ENDIF
        \ENDIF
        \IF{ $\mathbf{t} \neq \mathbf{h}$ }
          \STATE $S \leftarrow S\bigcup W_i$  
        \ENDIF
     \FOR{each vertex $\mathbf{v} \in S$}
       \STATE $H_j \leftarrow$ instantiate a new branch  
     \STATE $\mathrm{thisEntry}.\mathbf{v}\_par\_b \leftarrow \mathrm{thisEntry}.\mathbf{v}\_par\_b \bigcup loc\_H_j$ // record parallel branches location of $\mathbf{v}$ in $H_i$
      \STATE $(\mathrm{Entry})\ \mathbf{v}\_seq, \mathbf{v}\_post\_d, \mathbf{v}\_par\_b \leftarrow 0,1,\O$
     \STATE $H_j \leftarrow H_j \bigcup \mathrm{thisEntry}.\mathbf{v}$
        \STATE $(\mathrm{Entry}) \ \mathbf{q}\_seq, \mathbf{q}\_post\_d, \mathbf{q}\_par\_b \leftarrow 1,0,\O$
      \STATE $H_j \leftarrow H_j \bigcup \mathrm{thisEntry}.\mathbf{q}$
     \ENDFOR
          
\end{algorithmic} 
\end{algorithm}

\begin{algorithm}[h]
\renewcommand{\thealgocf}{\ref{subsec:bitgraph}.3}
\caption{$\mathsf{Bitgraph.Search}(\mathbf{q}, \theta; H, (\mathbf{ev}, loc\_H_a))$} 
\label{alg:bitgraph.search}
 \begin{algorithmic}[1]
\renewcommand{\algorithmicrequire}{\textbf{Input}:} 
      \REQUIRE a query $\mathbf{q}$, maximum nearest neighbor number $\theta$; 
      a bitgraph $H$, with an enter vertex $\mathbf{ev}$ and its   branch location $loc\_H_a$ (i.e., it is not necessarily the head entry).   
\renewcommand{\algorithmicrequire}{\textbf{Output}:} 
      \REQUIRE nearest neighbor vertices to a query $\mathbf{q}$.
\renewcommand{\algorithmicrequire}{\textbf{Search Procedure on $H$} ---} 
      \REQUIRE
      \STATE $E\leftarrow (\mathbf{ev}, loc\_H_a)$ // set of evaluated vertices and their branch locations
      \STATE $C\leftarrow (\mathbf{ev}, loc\_H_a)$ // queue of candidates and their branch locations
      \STATE $W\leftarrow (\mathbf{ev}, loc\_H_a)$  // queue of found nearest neighbors
      \WHILE{$|C|>0$}
      \STATE $(\mathbf{c}, loc\_H_i) \leftarrow$ extract nearest element from $C$  
      \STATE $\mathbf{f} \leftarrow$  get furthest element from $W$ to $\mathbf{q}$
      \IF{$\mathsf{Distance}(\mathbf{c}, \mathbf{q}) > \mathsf{Distance} (\mathbf{f}, \mathbf{q})$}
    \STATE \textbf{break}
      \ENDIF
      \textcolor{blue}{ \STATE // \textit{at-hand-detour} rule 
       \WHILE{$\mathbf{c}\_post\_d =0$}
      \STATE remove nearest element from $C$ 
      \STATE  $(\mathbf{c}, loc\_H_i) \leftarrow $ extract nearest element from $C$ 
      \ENDWHILE   }
     \FOR{each $(\mathbf{v}, loc\_H_j) \in \mathsf{Bitgraph.HoneyCombNeighbors}(\mathbf{c}, loc\_H_i)$   }
      \IF{($\mathbf{v}, loc\_H_j) \notin E$}
         \STATE $E \leftarrow E \cup (\mathbf{v}, loc\_H_j)$ 
      \STATE $\mathbf{f} \leftarrow$ get furthest element from $W$ 
      \STATE if $\mathsf{Distance}(\mathbf{v}, \mathbf{q}) < \mathsf{Distance}(\mathbf{f}, \mathbf{q})$
      \STATE $C\leftarrow C \cup (\mathbf{v}, loc\_H_j)$
       \STATE $W\leftarrow W \cup (\mathbf{v}, loc\_H_j)$
       \IF{$|W|> \theta$}
       \STATE remove furthest element from $W$
       \ENDIF
      \ENDIF
      \ENDFOR  
      \ENDWHILE
      \RETURN $W$
\end{algorithmic} 
\end{algorithm}

We also provide illustrative examples to clarify the functionalities.

\begin{figure}[h]
\includegraphics[scale=0.47]{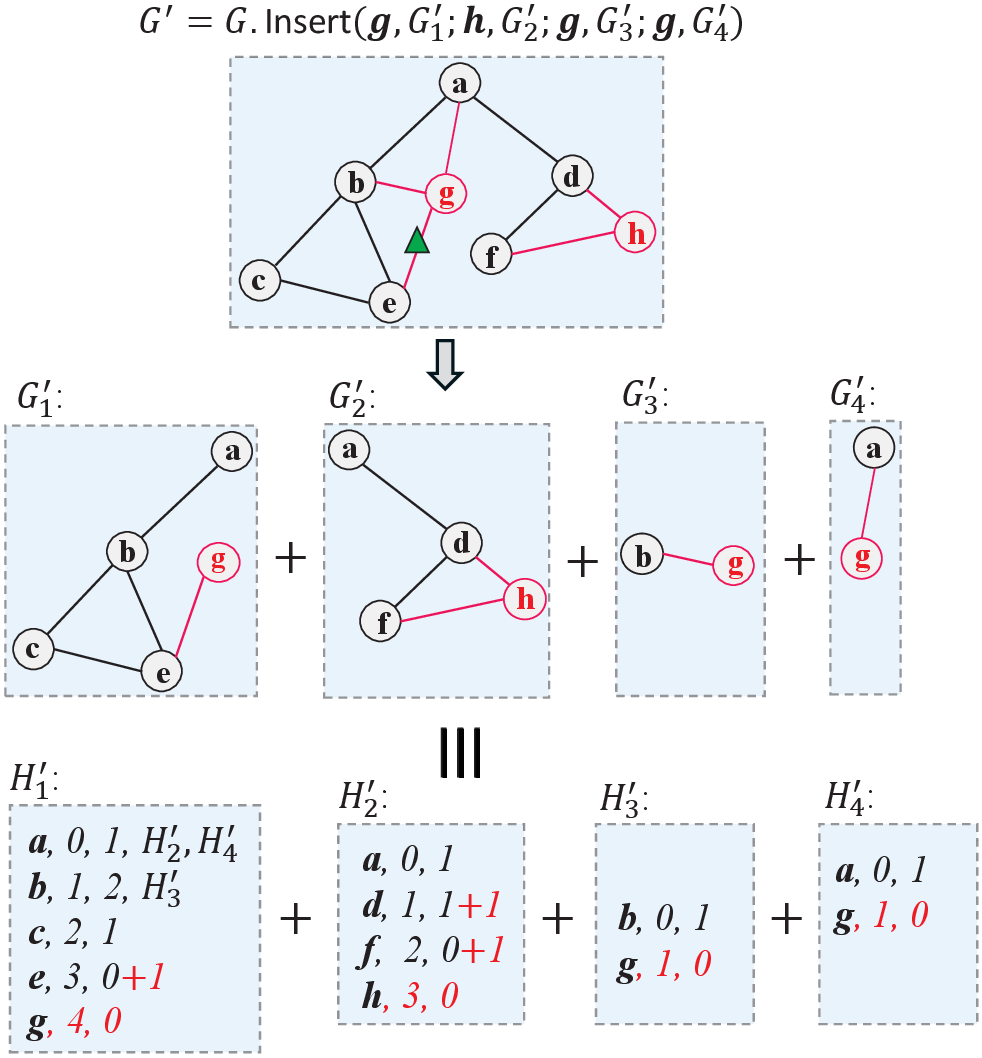}
\caption{A bigraph insert/search example}
\label{fig:exg_bitgraph_func}
\end{figure}

\textit{Example with Insert/Search Functionality.} Figure \ref{fig:exg_bitgraph_func} shows how values change when vertex $\mathbf{g}$ is inserted into branches $H'_1$, $H'_3$, $H'_4$ and vertex $\mathbf{h}$ into branch $H'_2$, where detailed insertion procedures is bypassed. We focus on the search process for a query vertex (represented by a green triangle). In the original graph structure, the search walk would proceed by entering vertex $\mathbf{a}$, analyzing $\mathbf{a}$'s neighbors, then examining $\mathbf{g}$'s neighbors to identify nearest neighbors (likely $\mathbf{g}$ and $\mathbf{e}$). However, in the bitgraph context, standard search protocols cannot establish a connection from $\mathbf{g}$ to $\mathbf{e}$ in this situation. Here, the \textit{at-hand-detour} function provides the solution by backtracking to vertex $\mathbf{b}$ and examining its neighborhood to successfully reach vertex $\mathbf{e}$.

\textsc{Correctness.} The walk isomorphism definition is derived from bitgraph isomorphism, providing the theoretical basis to validate that substituting bitgraphs in the search for graphs achieves correctness (i.e., identical search results) with acceptable deviation.

\begin{deduction}[\textit{Walk Isomorphism}] For any undirected graph $G$ and its isomorphic bitgraph $H$, there always exists at least one isomorphic walk in $H$  that covers any given walk in  $G$, regardless of which vertex in $H$ is selected to split branches. That is,  there exists 
    \begin{equation}
        f: \mathrm{walk}(G) \rightarrow \mathrm{walk}(H),
    \end{equation}
\end{deduction}
\noindent
such that the set of vertices in a walk of $G$ forms a subset of the vertices in the corresponding walk of $H$.

\textsc{Complexity Analysis.} 
Following Sec. \ref{subsec:necessity_bitgraph}, we quantify bitgraph sharing complexity by considering the minimum requirement for collective computation: all parties must together hold shares of at least all branch's vertices. The complexity metric becomes the sum of entries across all branches in a bitgraph, which characterizes the complexity for both sharing a bitgraph and searching distributed shares.

\begin{deduction}
For any bitgraph $H$ and undirected graph $G=\{V, E\}$,  $\{f_i: G_i \rightarrow H_i| G\overset{\Gamma}{=}\underset{i}{\bigcup}(G_i), H=\underset{i}{\bigcup}(H_i) \}$, assuming  $\mathbf{v}\in \mathbb{R}^d$ for any vertex $\mathbf{v}\in V$, when sharing $H$ among $n$ parties using $\mathsf{SS.Share}_n^t$ and \textit{iff} the vertices set of $H$ being shared,  the computational complexity  is:
\begin{equation}
    \begin{split}
        O(\mathsf{SS.Share}_n^t(H)) & = O(\mathsf{SS.Share}_n^t(\bar{V})) \\ 
        & = d(|V|+|H|) \cdot n \cdot O(p^t)\\
        & \cong d|V| \cdot n \cdot O(p^t),
   \end{split}
\end{equation}
\noindent
where  $|H|$ is the number of branches, corresponding to the count of split vertices; and $\bar{V}$ and $V$ are vertices (including repeated vertices) contained in $H$ and distinct vertices respectively.
In practical graph applications, this value is typically treated as a constant in the structure, with components like edges capturing meaningful relationships such as distances.
\end{deduction}

\begin{deduction}
\label{dec:O(search_on_shares_H)}
Let $c$ denote the number of vertices in a searching walk over any undirected graph $G=\{V, E\}$ with its isomorphic bitgraph $H$,  $\{f_i: G_i \rightarrow H_i| G\overset{\Gamma}{=}\underset{i}{\bigcup}(G_i), H=\underset{i}{\bigcup}(H_i) \}$, and $\mathbf{v}\in \mathbb{R}^d$ for any vertex $\mathbf{v}\in V$. Then, when sharing $H$ among $n$ parties using $\mathsf{SS.Share}_n^t$ and \textit{iff}  its vertices set $\bar{V}$ being shared, the computational complexity of a search operation over $H$'s shares is:
\begin{equation}
     \begin{split}
   & \quad    O(\mathsf{Search\_on\_Shares}(\{u, H_u\}_{u \in \mathcal{U}})) \\ 
   &= O(\mathsf{Search\_on\_Shares}(\{u, \bar{V}_u\}_{u \in \mathcal{U}}))  \\
   &
   = d(c+a) \cdot n \cdot O(p^t) \cdot O(\mathsf{AC}) \\
     &  \cong dc \cdot n \cdot O(p^t) \cdot O(\mathsf{AC}),   
     \end{split}
\end{equation}
\noindent
where $a$ is the number of additional vertices introduced by the \textit{at-hand-detour} function.
\end{deduction}

\section{A SP-A$^2$NN Search Scheme Based on HNSW} \label{sec:sp-a2nn}

\subsection{Construction} \label{subsec:sp-a2nn_construction}
Prior to presenting the complete algorithms, this section covers two key aspects: the storage structure detailing stored index components, data repository, and their inter-connections; and the search/update intuition for operating shared bitgraphs that are organized in the HNSW-indexing pattern, which explains the contents of bitgraph storage units that enable direct query token matching for SP-A$^2$NN searches. 
We then briefly discuss potential security enhancements to this work. Finally, we analyze the scheme in terms of complexity, parameter-related correctness, and security guarantees.

\textbf{Structure} \textit{-- Encrypted Database $C\text{-}EDB$}. 
The SP-A$^2$NN search scheme adheres to the conceptual structure defined in Sec \ref{subsec:dynamic SST}, Formula (\ref{equa:C_EDB}), organizing its database into separate index and data repositories. Like HNSW's multilayer graph index, which arranges vertices in layers of increasing density from sparse at the top to dense at the bottom, the encrypted collaborative index  in SP-A$^2$NN also consists of multiple hierarchical layers (i.e., $C\text{-}E\mathcal{I}\text{-}l$). Similarly, the bottom layer serves as the data repository containing vectors. The fundamental difference lies in that each layer uses an encrypted bitgraph (i.e., $C\text{-}EH\text{-}l$) in a shared pattern instead of an undirected graph employed in HNSW. We have
\begin{equation}
\label{equa:sp_a2nn_C_EDB}
\begin{split}
    C\text{-}EDB & = C\text{-}E\mathcal{I}+  C\text{-}ED \\ & = \sum_{1}^{L} C\text{-}E\mathcal{I}\text{-}l+  C\text{-}ED \\
    & = \sum_{1}^{L}  C\text{-}EH\text{-}l +  C\text{-}ED.               
\end{split}
\end{equation}

\textbf{Search/Update Intuition} \textit{-- From Bitgraph to Shared Bitgraph in HNSW Organization.}  
Searching in SP-A$^2$NN's collaborative encrypted index (i.e., $C\text{-}E\mathcal{I}$) proceeds layer-wise from top to bottom in the same manner as HNSW, finding query vector nearest neighbors in each layer until retrieving all bottom-layer data vectors. In a similar manner, insertion of a new element into $C\text{-}E\mathcal{I}$ relies on the search algorithm to preliminarily identify nearest neighbors from top to bottom, establishing the optimal placement for the new element. The overall search/insert architecture is shown in Algorithm \ref{alg:alg:sp_a2nn.insert} $\&$ \ref{alg:sp_a2nn.search}.

Let us now focus on searching within each layer where a bitgraph resides. The problem becomes clear given that Sec \ref{sec:technical_intuitions} already addresses searching and inserting vertices in bitgraphs. The operation of searching or inserting vertices within a layer (e.g, $C\text{-}E\mathcal{I}\text{-}l$) is equivalent to operating on a shared bitgraph (e.g., $C\text{-}EH\text{-}l$) where vertices are distributed among participating parties.
This architectural choice emphasizes data protection at the expense of leaving graph connectivity unencrypted in terms of \textit{access patterns} at every party's view, where connectivity means partial edges. Any party can easily recognize whether two vertices are adjacent by observing their storage locations and sequences, but cannot determine what the vertex content actually represents.

In particular, search operations implement a shared version that retains the core search logic from $\mathsf{Bitgraph.Search}$ in Sec \ref{sec:technical_intuitions},  while this algorithm itself is an unshared, unprotected plaintext version for identifying neighbors of a query vector within the bitgraph network. 
Direct query token matching in the shared scheme is implemented via share-based calculations given the distributed nature of vertices across parties. 
The algorithm of $\mathsf{Search\_Layer}$ is detailed in Algorithm \ref{alg:search_layer}. Insert operations $\mathsf{Insert\_Layer}$  (Alg \ref{alg:insert_layer}) follow a similar approach, mirroring the $\mathsf{Bitgraph.Insert}$ algorithm.

\textbf{Optimization, Detail and Discussion.} 
During search execution, a small adjustment for efficiency is made to the way $\mathsf{Bitgraph.Search}$ (Alg \ref{alg:bitgraph.search}) tracks evaluated vertices. Rather than maintaining the set and determining vertex membership status, this is replaced with bit-based information recording. That is, a standard bitgraph vertex quad $(\mathbf{v}, \mathbf{v}\_seq, \mathbf{v}\_post\_d, \mathbf{v}\_par\_b)$ is extended by adding a visit bit $\mathbf{v}\_e$ that marks whether a vertex has been accessed during search, preventing repeated vertex evaluations.
The complete algorithms are provided in Appendix \ref{apx:sp_a2nn_algorithms}. (We omit the SetUp algorithms for parties agreeing on keys and the state of each execution.)

Note that this version only covers insert scenarios, with delete methods left out of scope. SP-A$^2$NN's local database setting makes privacy concerns from dynamic updates, such as \textit{forward/backward privacy}, acceptable. Consequently, flexible deletion processes are possible, such as having all parties collectively mark a unit as deleted. However, when considering whether updating a unit belonging to one party might leak information about that unit's content to other parties, this introduces a distinct issue requiring further study.
Beyond the current scope, adding consistency verification mechanisms for this work's solutions can extend the scheme to provide protection against active adversaries.

\begin{figure*}[t]
\renewcommand{\baselinestretch}{1.1}
\small
\centering
  \fboxsep=2pt
  \fbox{
  \begin{minipage}[t]{0.32 \dimexpr\linewidth-2\fboxsep-40\fboxrule}
  \begin{algorithmic}[1]
 \renewcommand{\algorithmicrequire}{\underline{$\mathsf{Setup}(1^\lambda, \sigma)$}}
 \REQUIRE 
    \STATE $sd_i \overset{\$}{\leftarrow}  \{0,1\}^\lambda$ allocate list $L$
    \STATE Initiate Counter $\sigma: c \leftarrow 0$
    \STATE $K_i \leftarrow  F_1(sd_1, c)$ 
    \STATE Add $K_i$ into list $L$ (in lex order)
    \STATE Output $K=(K_i,  \sigma)$
  \end{algorithmic}
  \end{minipage}

\begin{minipage}[t]{0.38 \dimexpr\linewidth-2\fboxsep-40\fboxrule}
  \begin{algorithmic}[1]
 \renewcommand{\algorithmicrequire}{\underline{$\mathsf{Insert}(K, \sigma, \mathbf{q}; C\text{-}EDB)$}}
 \REQUIRE 
    \STATE (\textit{party} $u$) $\{\mathbf{q}_u\}_\mathcal{U} \leftarrow \mathsf{Enc}(\mathbf{q}, K_1)$ 
    \vspace{8pt}
     \STATE Set \par
     $C\text{-}E\mathcal{I} \leftarrow $  \par
     \quad  $C\text{-}E\mathcal{I}.\mathsf{Add}(\mathcal{I}\text{-}{hnsw}\text{-}{bitg}: \{ loc(\mathbf{q}_u)\}_\mathcal{U}, $ \par
     \quad  $ \mathbf{q}\_seq, \mathbf{q}\_post\_d, \mathbf{q}\_par\_b; \mathbf{q}\_e)$; \par
     $C\text{-}ED \leftarrow C\text{-}ED.\mathsf{Add}(\{\mathbf{q}_u\}_\mathcal{U}, $\par
     \quad $\mathbf{q}\_seq, \mathbf{q}\_post\_d, \mathbf{q}\_par\_b; \mathbf{q}\_e)$; \par
      $\sigma: c$++
     \STATE Output \par $C\text{-}EDB= (C\text{-}E\mathcal{I}, C\text{-}ED, \sigma)$
  \end{algorithmic}
  \end{minipage} 

  \begin{minipage}[t]{0.36 \dimexpr\linewidth-2\fboxsep-40\fboxrule}
  \begin{algorithmic}[1]
 \renewcommand{\algorithmicrequire}{\underline{$\mathsf{Search}(K, \sigma, \mathbf{q}; C\text{-}EDB)$}}
 \REQUIRE 
    \STATE (\textit{party} $v$) $\{\mathbf{q}_u\}_\mathcal{U} \leftarrow \mathsf{Enc}(\mathbf{q}, K_2)$ 
      \vspace{8pt}
    \STATE On input \par
    $\{\mathbf{q}\} \leftarrow v: \mathsf{Enc}(\mathbf{q})$
    \\$C\text{-}EDB = (C\text{-}E\mathcal{I}, C\text{-}ED, \sigma)$
    \STATE For $c=0$ until $\mathsf{SP\text{-}A^2NN}$(HNSW-Bitgraph index) return $\perp$, \par
    $\{\mathbf{v}_u\}_\mathcal{U}  \leftarrow $ \par \quad \q  $\mathsf{SP\text{-}A^2NN}(C\text{-}E\mathcal{I}; C\text{-}ED, \{\mathbf{q}\}_\mathcal{U}, F_2)$
   \STATE $\mathbf{v} \leftarrow \mathsf{Dec}(\{\mathbf{v}\}_\mathcal{U}, K)$
   \STATE Output $\mathbf{v}$
  \end{algorithmic}
  \end{minipage} }
  
  \caption{Real Scheme $\Pi_{SS}^{\mathcal{I}\text{-}hnsw\text{-}{bitg}}$}
    \label{fig:real_scheme}
\end{figure*}

\subsection{Analysis} 
\label{subsec:sp-a2nn_analysis}

\textsc{Complexity Analysis.} 
The complexity for a SP-A$^2$NN search  can be simplified to the complexity of searching on shares of a bitgraph (Deduction \ref{dec:O(search_on_shares_H)}) by the following reductions.

\begin{deduction} 
According to the component structure in Formula (\ref{equa:sp_a2nn_C_EDB}) of the multilayer $C\text{-}EDB$, the computational complexity of SP-A$^2$NN search can be decomposed into the complexity sum of searching each individual layer. Generally, we focus on the $l^{th}$ layer's encrypted bitgraph  $C\text{-}EH\text{-}l$ in a shared pattern, which is constrained to sharing only vertices and thus adheres to the complexity pattern of $\mathsf{Search\_on\_Shares}$. 

Let $\bar{V}$  represent the set of vertices actually traversed in a search walk over $C\text{-}EH\text{-}l$. This set's size is formulated using three parameters: $c$
is the vertex count in a search walk over the original HNSW graph that is converted from its isomorphic bitgraph ($H\text{-}l$), which quantifies the complexity of standard HNSW search; $a$ is the count of additional vertices triggered by \textit{at-hand-detour} functions; and $o$ is the number of deviated vertices incurred by \textit{honeycomb-neighbors} tracing walks compared to the original HNSW walks. Thus, the computational complexity of SP-A$^2$NN search compared to reference HNSW search is: 
\begin{equation}
    \begin{split}
       & \quad  O(\mathsf{SP\text{-}A^2NN.Search}(\{\mathbf{q}\}_\mathcal{U};C\text{-}EDB)) \\
        & = 
        O(\mathsf{SP\text{-}A^2NN.Search\_Layer}(\{\mathbf{q}\}_\mathcal{U}; \sum_{1}^{L}C\text{-}EH\text{-}l, C\text{-}ED)) \\
        & =  (L+1) \cdot O(\mathsf{Search\_on\_Shares}(\{u, \bar{V}_u\}_{u \in \mathcal{U}})))  \\
        & = (L+1) \cdot d(c+a+o) \cdot n \cdot O(p^t) \cdot O(\mathsf{AC})\\
        & \cong (L+1) \cdot dc \cdot n \cdot O(p^t) \cdot O(\mathsf{AC})\\
        & = O(\mathsf{HNSW.Search}(\mathbf{q};DB))\cdot n \cdot O(p^t) \cdot O(\mathsf{AC})
   \end{split}
\end{equation}
\noindent

\end{deduction}

Observe that SP-A$^2$NN search introduces only the additional overhead of computing distance comparisons via arithmetic circuits, relative to standard HNSW search on unencrypted data.

\textsc{Correctness and Security Analysis.}
We validate the correctness and security of a SP-$^2$ANN search scheme via a real instantiated construction as follows.  

\textit{Tunable Parameters Impact on Correctness.} In the bitgraph structure, the same vertex may be located in multiple branches, resulting in duplicate vertices from different branches potentially being recorded in the queue $W$, which dynamically maintains search results during the search operation. 
Additionally, the search results contain additional deviation vertices compared to the original HNSW results, since the actual search walks traverse $(a+o)$ additional vertices. Despite occasionally missing some vertices relative to the $c$ vertices in the original search walks, the scheme design guarantees tracing the original paths as faithfully as possible. We view this deviation as quite acceptable since, in real-world scenarios,  the search results in queue $W$ act as candidate sets, with final elements selected through closest-first or heuristic selection methods. Thus, in real applications,  the queue size $|W|$ (i.e., $\theta$) can be tuned to a relatively large range to avoid having repeated vertex positions negatively impact the expected search results.
The impact of this part is considered and incorporated when validating correctness of Theorem \ref{theo:Pi_correct}.

\textbf{Real SP-A$^2$NN-Instantiated Construction.}
Let a $(t,n)$-threshold secret sharing configuration $SS$ serve as an encryption scheme of $(\mathsf{Enc}, \mathsf{Dec})$, and $\mathcal{I}\text{-}hnsw\text{-}bitg$ be the bitgraph-based  HNSW index to organize $C\text{-}EDB$. $F_1$ and $F_2$ are the same as in $\Pi_{SS}$.
We have our real construction $\Pi_{SS}^{\mathcal{I}\text{-}hnsw\text{-}bitg}$ in Fig \ref{fig:real_scheme}.

\begin{theorem}  
\label{theo:Pi_correct}
A real scheme $\Pi^{\mathcal{I}\text{-}hnsw\text{-}bitg}_{SS}$ is $\Delta(\rho)$-correct \textit{iff} the reduction from $\Pi_{SS}^{\mathcal{I}\text{-}hnsw\text{-}bitg}$ to $\Pi_{SS}^{\mathcal{I}\text{-}hnsw}$ \textit{w.r.t correctness} is $\Delta(\Pi_{SS}^{\mathcal{I}\text{-}hnsw\text{-}bitg},\Pi_{SS}^{\mathcal{I}\text{-}hnsw})$-correct.
\end{theorem}

The proof for Theorem \ref{theo:Pi_correct} is omitted, while the impact from the tunable parameters is used to measure $\Delta(\rho)$, and we consider this impact on deviation allowable.

\begin{theorem}
\label{theo:Pi_secure}
A real scheme $\Pi^{\mathcal{I}\text{-}hnsw\text{-}bitg}_{SS}$ is  $\mathcal{L}(\epsilon)$-secure \textit{iff} the reduction from $\Pi_{SS}^{\mathcal{I}\text{-}hnsw\text{-}bitg}$ to $\Pi_{SS}^{\mathcal{I}\text{-}hnsw}$ \textit{w.r.t security}  and \textit{w.r.t leakage} is $\mathcal{L}(\Pi_{SS}^{\mathcal{I}\text{-}hnsw\text{-}bitg},\Pi_{SS}^{\mathcal{I}\text{-}hnsw})$-secure.
\end{theorem}
The proof for Theorem \ref{theo:Pi_secure} is in Appendix \ref{apx_proof:Pi to Pi_m}.

\clearpage
\bibliographystyle{unsrt}
%\bibliography{source.bib}

\begin{thebibliography}{1}

\bibitem{DBLP:conf/nips/LewisPPPKGKLYR020}
Patrick Lewis, Ethan Perez, Aleksandra Piktus, Fabio Petroni, Vladimir Karpukhin, Naman Goyal, Heinrich K{\"{u}}ttler, Mike Lewis, Wen{-}tau Yih, Tim Rockt{\"{a}}schel, Sebastian Riedel, and Douwe Kiela.
\newblock Retrieval-augmented generation for knowledge-intensive {NLP} tasks.
\newblock In {\em Advances in Neural Information Processing Systems 33: Annual Conference on Neural Information Processing Systems 2020, NeurIPS 2020, December 6-12, 2020, virtual}, 2020.

\bibitem{DBLP:journals/pami/MalkovY20}
Yury~A. Malkov and Dmitry~A. Yashunin.
\newblock Efficient and robust approximate nearest neighbor search using hierarchical navigable small world graphs.
\newblock {\em {IEEE} Trans. Pattern Anal. Mach. Intell.}, 42(4):824--836, 2020.

\bibitem{DBLP:journals/dam/Maurer06}
Ueli~M. Maurer.
\newblock Secure multi-party computation made simple.
\newblock {\em Discret. Appl. Math.}, 154(2):370--381, 2006.

\bibitem{DBLP:conf/sp/SongWP00}
Dawn~Xiaodong Song, David~A. Wagner, and Adrian Perrig.
\newblock Practical techniques for searches on encrypted data.
\newblock In {\em 2000 {IEEE} Symposium on Security and Privacy, Berkeley, California, USA, May 14-17, 2000}, pages 44--55. {IEEE} Computer Society, 2000.

\bibitem{DBLP:conf/sp/Servan-Schreiber22}
Sacha Servan{-}Schreiber, Simon Langowski, and Srinivas Devadas.
\newblock Private approximate nearest neighbor search with sublinear communication.
\newblock In {\em 43rd {IEEE} Symposium on Security and Privacy, {SP} 2022, San Francisco, CA, USA, May 22-26, 2022}, pages 911--929. {IEEE}, 2022.

\bibitem{Skip_list.wiki}
Wikipedia.
\newblock Skip list.
\newblock \url{https://en.wikipedia.org/wiki/Skip_list}.

\bibitem{DBLP:journals/cacm/Shamir79}
Adi Shamir.
\newblock How to share a secret.
\newblock {\em Commun. {ACM}}, 22(11):612--613, 1979.

\bibitem{DBLP:conf/ccs/BostMO17}
Rapha{\"{e}}l Bost, Brice Minaud, and Olga Ohrimenko.
\newblock Forward and backward private searchable encryption from constrained cryptographic primitives.
\newblock In {\em Proceedings of the 2017 {ACM} {SIGSAC} Conference on Computer and Communications Security, {CCS} 2017, Dallas, TX, USA, October 30 - November 03, 2017}, pages 1465--1482. {ACM}, 2017.

\bibitem{Graph_theory.wiki}
Wikipedia.
\newblock Graph theory.
\newblock \url{https://en.wikipedia.org/wiki/Graph_theory}.

\end{thebibliography}

\begin{appendices}

 \begin{figure*}[h]
 \centering
\includegraphics[scale=0.51]{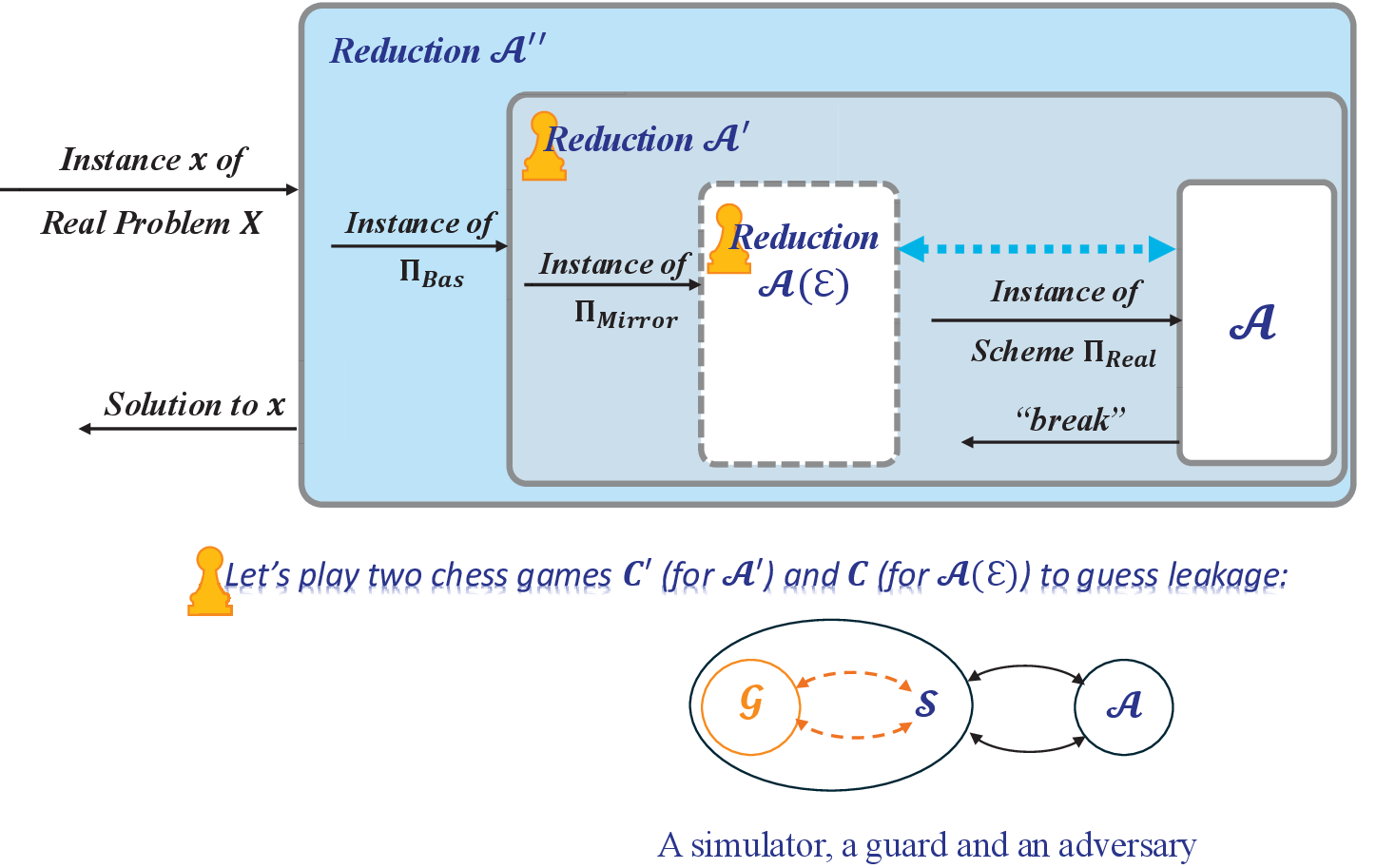}
\caption{A high-level overview of reduction framework for privacy analysis}
\label{fig:logical_frame}
\end{figure*}

\newpage
\section{A Leakage-Guessing Proof System for Privacy Analysis} 

\subsection{\textbf{Logical Reduction Framework}}
\label{apx:framework}

\subsection{Proof -- \textit{Reduction from  $\Pi$ 
 to $\Pi_{Bas}$}}

\subsubsection{Proof for Theorem \ref{theo:Pi_ss to ss} -- \textit{from $\Pi_{Bas}$ to $SS$}}
\label{apx_proof:Pi_bas to ss}

This part is assumed validated.

\subsubsection{Proof for Theorem \ref{theo:Pi_ss_hnsw to Pi_ss} -- \textit{from $\Pi_M$ to $\Pi_{Bas}$}} \quad
\label{apx_proof:Pi_m to Pi_bas}

\textsc{Settings:} We introduce a chess-play game aided by an oracle definition to complete the secruity proof. 

\begin{definition}[Oracle $\mathcal{F}$] The oracle $\mathcal{F}$ is defined as:
\begin{equation}
    \mathcal{F}: (x, f) \rightarrow \{f(x,y)  : y \in Connected(x)\}
\end{equation}
where $x\in X$ is the query element, the oracle returns the set of function values over all elements connected to
$x$, and $f: X \times X \rightarrow 
\{0,1\}$ is the underlying link existence function.
\end{definition}

\begin{proposition}[Chess Game $\mathcal{C}'$]
\label{apx_game: mirror game}
Let $\mathcal{C}$ be a oracle-$\mathcal{F}$-aided interaction-based protocol between simulators $S_1$, $S_2$, a guard $G$ and an adversary $\mathcal{A}$ to identify the related  information between two ciphers encrypted by two different schemes from an identical message.
\vspace{10pt}

\textbf{Init:} $S_2$ runs $\Pi_{SS}^{\mathcal{I}\text{-}{hnsw}}.\mathsf{Init}$, $\Pi_{SS}^{\mathcal{I}\text{-}{hnsw}}.\mathsf{Insert}$ in Fig. \ref{fig:mirror_scheme} and  $S_1$ runs $\Pi_{SS}.\mathsf{Init}$, $\Pi_{SS}.\mathsf{Insert}$ in Fig. \ref{fig:basic_scheme}, both in a completed execution state.

\textbf{Chess-Play:} \textsc{[$\mathcal{A}$ query $S$]} $S_1$ takes any cipher element $\mathbf{e}$ queried from $\mathcal{A}$ and outputs a result $(\mathbf{e}, C\text{-}ED;  C\text{-}E\mathcal{I}_1)$ to $\mathcal{A}$. Following the same pattern, $S_2$ outputs  $(\mathbf{e}, C\text{-}ED;  C\text{-}E\mathcal{I}_2)$ to $\mathcal{A}$.

\textsc{[$S$ request $G$]} Both $S_1, S_2$ backup their outputs to $\mathcal{A}$ and hand them over to $G$. 
For $S_1$, $G$ takes over $S_1$'s $\Pi_{SS}.\mathsf{Init}$ to decode $\mathbf{e}$ from $C\text{-}ED$, denoted as a message $\mathbf{m\_e}$, and reverses $C\text{-}E\mathcal{I}_1$ to $C\text{-}\mathcal{I}_1$.  Then $G$ invokes oracle-$\mathcal{F}$ with input $(\mathbf{m\_e}, C\text{-}\mathcal{I}_1: \cup f)$ to extract the complete adjacency information linking $\mathbf{m\_e}$ to all related elements,  yielding a set of elements $\{\mathbf{m\_e_1}\}$. (Here, $f$ contains the position and pointer structure of each element in $C\text{-}\mathcal{I}_1$, while $\cup f$ captures all structures in $C\text{-}\mathcal{I}_1$.) $G$ decodes this set to $\{\mathbf{e_1}\}$ via keys in $\Pi_{SS}.\mathsf{Init}$ and outputs a pair. The pair $(\mathbf{e}, \{\mathbf{e_1}\}; f_c)$ represents that there exists a connection from element $\mathbf{e}$ to elements in the set $\{\mathbf{e_1}\}$ within $C\text{-}E\mathcal{I}_1$, where the connection is formulated with $f_c:=\{loc(\mathbf{e_1})\}$ according to the execution in Fig \ref{fig:basic_scheme} (Line 2).  

Analogously, for $S_2$, $G$ outputs $(\mathbf{e}, \{\mathbf{e_2}\}; f_c)$ that represents there exists a connection from $\mathbf{e}$ to $\{\mathbf{e_2}\}$ in $C\text{-}E\mathcal{I}_2$, where the connection is formulated with $f_c:=\{\mathcal{I}\text{-}hnsw: loc(\mathbf{e_2})\}$ according to the execution in Fig \ref{fig:mirror_scheme} (Line 2).  

\textsc{[$G$ response $S$]} $G$ outputs the response to $S$.

\textsc{[$S$ answer $\mathcal{A}$]}  $S$ answers the outputs to $\mathcal{A}$.

\textbf{Bonus (Leakage): } What $\mathcal{A}$ obtains defines the bonus he wins in the game. 
\end{proposition}

\vspace{10pt}

\textsc{Security Proof:} The proof follows a two-step reduction pattern: first, we verify that the leakage reduction for
$\mathcal{L}(\Pi_{SS}^{\mathcal{I}\text{-}hnsw},\Pi_{SS})$-security holds, then we show that complete equivalence reduces from $\Pi_{SS}^{\mathcal{I}\text{-}hnsw}$ to $\Pi_{SS}$ (i.e., security of all practically meaningful encrypted data, e.g., $C\textit{-}ED$).

\textit{Step-1: Leakage Reduction.} To extract the leakage incurred when reducing from
$\Pi^{\mathcal{I}\text{-}hnsw}_{SS}$ to $\Pi_{SS}$, we begin with several deduction sketches as: 

\begin{apx-deduction}
\label{apx_dec:cEDB_leak}
 The leakage between $(\Pi^{\mathcal{I}\text{-}hnsw}_{SS}, \Pi_{SS})$ reduces to data of $C\text{-}EDB$'s leakage
between $(\Pi^{\mathcal{I}\text{-}hnsw}_{SS}, \Pi_{SS})$.
\end{apx-deduction}

\begin{apx-deduction}
\label{apx_dec:cEI_leak}
The $C\text{-}EDB$'s leakage
between ($\Pi^{\mathcal{I}\text{-}hnsw}_{SS}$, $ \Pi_{SS}$) reduces to leakage
between ($\Pi^{\mathcal{I}\text{-}hnsw}_{SS}$'s $C\text{-}E\mathcal{I}$, $\Pi_{SS}$'s $C\text{-}E\mathcal{I}$) \textit{if} complete equivalence holds in ($\Pi^{\mathcal{I}\text{-}hnsw}_{SS}$'s $C\text{-}ED$, $\Pi_{SS}$'s $C\text{-}ED$).
\end{apx-deduction}

\begin{apx-deduction}
\label{apx_dec:fulldata_for_cEI_leak}
The leakage
between ($\Pi^{\mathcal{I}\text{-}hnsw}_{SS}$'s $C\text{-}E\mathcal{I}$, $\Pi_{SS}$'s $C\text{-}E\mathcal{I}$) equals the total leakage of all elements in $C\text{-}ED$ that occur in both $C\text{-}E\mathcal{I}$s.
\end{apx-deduction}

\begin{apx-deduction}
\label{apx_dec:sindata_for_cEI_leak}
The leakage of any element in $C\text{-}ED$ that occur in both $C\text{-}E\mathcal{I}$s can be calculated via a privacy-guessing game.
\end{apx-deduction}

To validate Deduction \ref{apx_dec:sindata_for_cEI_leak}, we construct a simulator $S'_0$ to simulate an individual party's view (e.g., party $i$), which invokes a leakage-guessing game called chess game $\mathcal{C}'$ (Proposition \ref{apx_game: mirror game})  for any element $\mathbf{e}$ of $C\text{-}ED$ as inputs. From a high-level respective, party $i$'s view captures the highest level of privilege, not only allowing complete observation of ciphers flowing both within/through party $i$'s territory, but also invoking Enc/Dec oracle for reversing any cipher to its message. This privilege is transferred to a guard role $G$ of $\mathcal{C}'$. We have

\begin{equation}  
\begin{split}
& \quad \mathcal{L}_i(\Pi_{SS}^{\mathcal{I}\text{-}hnsw}, \Pi_{SS}) \\  
& =  \mathcal{L}_i(\Pi_{SS}^{\mathcal{I}\text{-}hnsw}: (C\text{-}E\mathcal{I}, C\text{-}ED),  \Pi_{SS}: (C\text{-}E\mathcal{I}, C\text{-}ED)) \\
& = \mathcal{L}_i(\Pi_{SS}^{\mathcal{I}\text{-}hnsw}: C\text{-}E\mathcal{I},  \Pi_{SS}: C\text{-}E\mathcal{I}; C\text{-}ED)
\\ & = 
\sum_\mathbf{e} \mathcal{L}_i (
C\text{-}E\mathcal{I}_2, C\text{-}E\mathcal{I}_1;  C\text{-}ED, \mathbf{e}) \ \textit{for} \ \mathbf{e}\in C\text{-}ED \\
  & = \sum_\mathbf{e}  S'_0.\mathcal{C}'(C\text{-}E\mathcal{I}_2, C\text{-}E\mathcal{I}_1;  C\text{-}ED, \mathbf{e})  \ \textit{for} \ \mathbf{e}\in C\text{-}ED
\end{split}
\end{equation}
where the equalities are justified as follows: the $1$st by deduction of \ref{apx_dec:cEDB_leak}, the 2nd by by deduction of \ref{apx_dec:cEI_leak}, the 3th by by deduction of \ref{apx_dec:fulldata_for_cEI_leak}, and the 4th by deduction of \ref{apx_dec:sindata_for_cEI_leak} and the followed deduction. (For $\mathbf{e}$'s format, taking Fig. \ref{fig:basic_scheme} as an example, $\mathbf{e}$ is 
a share of $\{\mathbf{q}_i\}_\mathcal{U}$ hold by party $i$.))

\textit{Step-2: complete $C\text{-}ED$ equivalence.} It can be observed that the validation of Deduction \ref{apx_dec:cEI_leak} relies on complete equivalence on ($\Pi^{\mathcal{I}\text{-}hnsw}_{SS}$'s $C\text{-}ED$, $\Pi_{SS}$'s $C\text{-}ED$). Examining the execution of inserting any element into 
$C\text{-}ED$ is identical in both Fig \ref{fig:basic_scheme} and Fig \ref{fig:mirror_scheme}, this complete equivalence holds.

\textsc{Correctness Baseline.} As in formula (\ref{equa:cor_Pi_m}) (Def \ref{def:sst_cor}), the correctness of mirror construction $\Pi^{\mathcal{I}\text{-}hnsw}_{SS}$ 
is the reference baseline under the problem $\Sigma$.

\textsc{Conclusion.} Back to Theorem \ref{theo:Pi_ss_hnsw to Pi_ss}, we have its $\mathcal{L}$-secure claim on $(\Pi_{M}, \Pi_\text{Bas})$ is hold as:
\begin{equation}
\begin{split} 
   & \quad \mathcal{L}(\Pi_{M}, \Pi_\text{Bas}) \\
   & = \sum_\mathbf{e}  (\mathbf{e}, \{\mathbf{e_2}\}; \mathcal{I}\text{-}{hnsw}:\{loc(\mathbf{e_2})\}) \overset{def}{-} (\mathbf{e}, \{\mathbf{e_1}\}; \{loc(\mathbf{e_1})\})  \\
   & = \sum_\mathbf{e}  (\mathbf{e}, \{\mathbf{e_2}\}; \mathcal{I}\text{-}{hnsw}:\{loc(\mathbf{e_2})\}) 
\end{split}
\end{equation} 
where the 2nd equality is justified by the independence of storage locations due to no indexing occurring in  $\mathbf{e}, \{\mathbf{e_1}\}; \{loc(\mathbf{e_1})\}$.

The correctness claim on $(\Pi_{M}, \Pi_\text{Bas})$ holds by referring to above correctness baseline.

\subsubsection{Proof for Theorem \ref{theo:Pi_ss to ss} -- \textit{from $\Pi$ to $\Pi_M$}} \quad
\label{apx_proof:Pi to Pi_m}

\textsc{Settings:}  A chess-play game $\mathcal{C}$ analogous to that in Appendix \ref{apx_proof:Pi_m to Pi_bas} is defined, where the difference lies in:  given $\mathcal{A}$'s query $\mathbf{e}$, during \textbf{Init}, $S_2$ runs 
$\Pi_{SS}^{\mathcal{I}\text{-}hnsw\text{-}bitg}$'s $\mathsf{Init}$ and $\mathsf{Insert}$ algorithms in Fig. \ref{fig:real_scheme}, and  $S_1$ runs $\Pi_{SS}^{\mathcal{I}\text{-}hnsw}$'s  $\mathsf{Init}, \mathsf{Insert}$ in Fig. \ref{fig:mirror_scheme}, with both achieving a completed execution state. The final output of the game $\mathcal{C}$ is $(\mathbf{e}, \{\mathbf{e_2}\}; f_{c2})$  and  $(\mathbf{e}, \{\mathbf{e_1}\}; f_{c_1})$, where $f_{c2}$ is 
 \\
$\{\mathcal{I}\text{-}hnsw\text{-}bitg: loc(\mathbf{e_2}), \mathbf{e_2}\_seq, \mathbf{e_2}\_post\_d, \mathbf{e_2}\_par\_b; \mathbf{e_2}\_e\}$ and 
$f_{c1}$ is $\{\mathcal{I}\text{-}hnsw: loc(\mathbf{e_1})\}$.

\vspace{10pt}
\textsc{Security Proof:} The proof also uses a two-step reduction pattern as in Appendix \ref{apx_proof:Pi_m to Pi_bas}: $\mathcal{L}$-security reduction and complete equivalence reduction (exclusively in terms of  $C\text{-}ED$) from $\Pi_{SS}^{\mathcal{I}\text{-}hnsw\text{-}bitg}$ to $\Pi_{SS}^{\mathcal{I}\text{-}hnsw}$.

\textit{Step-1: Leakage Reduction.} Adopting the same framework for reductions  (Apx \ref{apx_proof:Pi_m to Pi_bas} Step-1), we let $S_0'$ simulate the view for an individual party $i$ and invoke $\mathcal{C}$, yielding
\begin{equation}  
\begin{split}
& \quad \mathcal{L}_i(\Pi_{SS}^{\mathcal{I}\text{-}hnsw\text{-}bitg}, \Pi_{SS}^{\mathcal{I}\text{-}hnsw}) \\  
& =  \mathcal{L}_i(\Pi_{SS}^{\mathcal{I}\text{-}hnsw\text{-}bitg}: (C\text{-}E\mathcal{I}, C\text{-}ED),  \Pi_{SS}^{\mathcal{I}\text{-}hnsw}: (C\text{-}E\mathcal{I}, C\text{-}ED)) \\
& = \mathcal{L}_i(\Pi_{SS}^{\mathcal{I}\text{-}hnsw\text{-}bitg}: (C\text{-}E\mathcal{I}),  \Pi_{SS}^{\mathcal{I}\text{-}hnsw}: (C\text{-}E\mathcal{I}: (C\text{-}E\mathcal{I}); C\text{-}ED)
\\ & = 
\sum_\mathbf{e} \mathcal{L}_i (
C\text{-}E\mathcal{I}_2, C\text{-}E\mathcal{I}_1;  C\text{-}ED, \mathbf{e}) \ \textit{for} \ \mathbf{e}\in C\text{-}ED  \\
& = \sum_\mathbf{e}  S_0.\mathcal{C}(C\text{-}E\mathcal{I}_2, C\text{-}E\mathcal{I}_1;  C\text{-}ED, \mathbf{e})  \ \textit{for} \ \mathbf{e}\in C\text{-}ED
\end{split}
\end{equation}

\textit{Step-2: complete $C\text{-}ED$ equivalence.} The equivalence on ($\Pi^{\mathcal{I}\text{-}hnsw}_{SS}$'s $C\text{-}ED$, $\Pi_{SS}$'s $C\text{-}ED$) is validated since 
the identical execution of inserting any element into 
$C\text{-}ED$ occurs in both Fig \ref{fig:real_scheme} and Fig \ref{fig:mirror_scheme}.

Back to Theorem \ref{theo:Pi_secure}, we have its $\mathcal{L}$-secure claim on $(\Pi, \Pi_M)$ is hold as:
\begin{equation}
\begin{split} 
   & \quad \mathcal{L}(\Pi, \Pi_M) \\
   & = \sum_\mathbf{e}  (\mathbf{e}, \{\mathbf{e_2}\}; f_{c2})  \overset{def}{-} (\mathbf{e}, \{\mathbf{e_1}\}; \{\mathcal{I}\text{-}{hnsw}:\{loc(\mathbf{e_2})\})  \\
   & = \sum_\mathbf{e}  (\mathbf{e}, \{\mathbf{e_2}\}; f_{c2}) 
\end{split}
\end{equation} 
where $f_{c2}$ is \\
$\{\mathcal{I}\text{-}hnsw\text{-}bitg: loc(\mathbf{e_2}), \mathbf{e_2}\_seq, \mathbf{e_2}\_post\_d, \mathbf{e_2}\_par\_b; \mathbf{e_2}\_e\}$.

Drawing upon the definition of privacy triplet (Def \ref{def:privacy triplet}), for any individual, randomly chosen data element $\mathbf{e}$ where its message is assumed to be learned, we can calculate 
the leakage exposure $\mathcal{L}(\epsilon)$ incurred by $\mathbf{e}$ on the multilayer-bitgraph organized $C\text{-}EDB$, specifically $\sum_1^L C\text{-}EH\text{-}l+C\text{-}ED$. Through a I-III trajectory, we have 

\begin{equation*}
 \mathcal{L}_I^{\mathcal{I}\text{-}hnsw\text{-}bitg}(\mathbf{e})= \frac{(L+1)\times (1+post\_d + par\_b)}{C\text{-}ED}.
\end{equation*}
where the first $1$ counts for the prior one of $\mathbf{e}$ in sequence of a layer.

\begin{equation*}
\mathcal{L}_{II}^{\mathcal{I}\text{-}hnsw\text{-}bitg}(\{\mathbf{e_2}\})=  \frac{\sum_{|\{\mathbf{e_2}\}|} (L+1)\times (1+post\_d + par\_b)}{C\text{-}ED}
\end{equation*}

Having traversed interfaces I, II, we can determine the number of elements that exhibit connections with $\mathbf{e}$.To further measure  this connection, based on knowledge of $\mathcal{L}(\Pi, \Pi_M)$ and public parameters, we can state

\begin{equation*}
\begin{split}
& \quad \quad \mathcal{L}_{III}^{D}(\mathbf{e}, \{\mathbf{e_2}\}) = \\ & \frac{\{\mathsf{Distance}(\theta) | \sum_{|\{\mathbf{e_2}\}|+1} (L+1)\times (1+post\_d + par\_b)\} }{C\text{-}ED} 
\end{split}
\end{equation*}
where $\mathsf{Distance}(\theta)$ measures similarity distance of two elements (i.e., vectors) given that we treat the query range threshold $\theta$ is public  (although we examine $\theta$ as the maximum number of neighbors a query contains in our work).

\textsc{Conclusion.} Back to Theorem \ref{theo:Pi_secure}, we have its $\mathcal{L}(\epsilon)$-secure claim on $(\Pi, \Pi_M)$ is hold as:

\begin{equation*}
\begin{split}
& \quad \quad \mathcal{L}(\epsilon)= \mathcal{L}_{III}^{D}(\mathbf{e}, \{\mathbf{e_2}\}) = \\ & \frac{\{\mathsf{Distance}(\theta) | \sum_{|\{\mathbf{e_2}\}|+1} (L+1)\times (1+post\_d + par\_b)\} }{C\text{-}ED}.
\end{split}
\end{equation*}

\section{SP-A$^2$NN Algorithms} \label{apx:sp_a2nn_algorithms}

\begin{algorithm}[h]
\renewcommand{\thealgocf}{\ref{apx:sp_a2nn_algorithms}.2}
\caption{$\mathsf{SP\text{-}A^2NN.Insert}(K_\mathsf{SS,AC}, \sigma, \{\mathbf{q}\}_{\mathcal{U}}, l'; \\
C\text{-}EDB)$}  
\label{alg:alg:sp_a2nn.insert}
 \begin{algorithmic}[1]
\renewcommand{\algorithmicrequire}{\textbf{Input}:} 
      \REQUIRE a new vector $\{\mathbf{q}\}_\mathcal{U}$ submitted by party $v$, this new element’s level $l'$; \\
    $C\text{-}EDB$:  multiple bitgraphs, and its an enter vector $\{\mathbf{ev}\}_{\mathcal{U}}$ shared from party $u$, which is  located in branch $H_a$ of top layer's bitgraph (i.e., the $L^{th}$ layer). \\
   \textit{(Locations of bitgraph, branches, and units are public parameters.)}
\renewcommand{\algorithmicrequire}{\textbf{Output}:} 
      \REQUIRE  
\renewcommand{\algorithmicrequire}{\textbf{Insert Procedure} ---} 
      \REQUIRE
      \STATE $(\{\mathbf{ev}\}_{\mathcal{U}}, loc\_H_a) \leftarrow$ get enter vector for $C\text{-}EDB$
       \FOR{$l \leftarrow L ... l'+1$}
      \STATE $\{W\}_\mathcal{U} \leftarrow \mathsf{Search\text{-}Layer}(\{\mathbf{q}\}_{\mathcal{U}}, \theta=1; $\\ \quad \quad \quad \quad  $ C\text{-}E\mathcal{I}\text{-}l, (\{\mathbf{ev}\}_{\mathcal{U}}, loc\_H_a))$
      \STATE $(\{\mathbf{ev}\}_{\mathcal{U}}, loc\_H_a) \leftarrow$ get first element from $\{W\}_\mathcal{U}$ 
      \ENDFOR
      \FOR{$l \leftarrow l' ... 0$}
      \STATE $\{W\}_\mathcal{U} \leftarrow \mathsf{Search\text{-}Layer}(\{\mathbf{q}\}_{\mathcal{U}}, \theta; $\\ \quad \quad \quad \quad  $ C\text{-}E\mathcal{I}\text{-}l, (\{\mathbf{ev}\}_{\mathcal{U}}, loc\_H_a))$
      \STATE $(\mathbf{ev}, loc\_H_a) \leftarrow$ get first element from $ \{W\}_\mathcal{U}$
      \ENDFOR      
\end{algorithmic} 
\end{algorithm}

\begin{algorithm}[h]
\renewcommand{\thealgocf}{\ref{apx:sp_a2nn_algorithms}.1}
\caption{
$\mathsf{Insert\text{-}Layer}(\{\mathbf{q}\}_\mathcal{U}$; \quad  \\ $C\text{-}E\mathcal{I}\text{-}l, (\{W_i\}_\mathcal{U}, loc\_H_i)$ )} 
\label{alg:insert_layer}
    \begin{algorithmic}[1]
\renewcommand{\algorithmicrequire}{\textbf{Clients Input}:}
    \REQUIRE  a new vector $\{\mathbf{q}\}_\mathcal{U}$ shared from party $v$; \\
    $C\text{-}E\mathcal{I}\text{-}l, (\{W_i\}_\mathcal{U}, loc\_H_i)$:  the $l^{th}$ layer's bitgraph, and $\{\mathbf{q}\}_\mathcal{U}$'s pre-positive vertices set $\{W_i\}_\mathcal{U}$ with an ordered sequence, its branch location  $loc\_H_i$ \\ (i.e., $\{W_i\}_\mathcal{U} \subseteq H_i$). 
\renewcommand{\algorithmicrequire}{\textbf{Clients Output}:}
    \REQUIRE
    update branch $H_i$ in $C\text{-}E\mathcal{I}\text{-}l$  after inserting $\mathbf{q}$.
\renewcommand{\algorithmicrequire}{\textbf{Insert Procedure on a Layer} ---} 
        \REQUIRE
\renewcommand{\algorithmicrequire}{\textit{all parties in  $\mathcal{U}$}:}     
  \REQUIRE
    \STATE $S \leftarrow \O$ // set of new split vertices
     \STATE $\{H_j: H_j \leftarrow \O \}$ // set of new branches produced from the split vertices in $H_i$
      \STATE $\{T\}_\mathcal{U} \leftarrow$ agree on set of vertices having continuous sequences from $\{W_i\}_\mathcal{U}$ in a back-to-front order
      \STATE $\{\mathbf{t}\}_\mathcal{U} \leftarrow$ get last element from $\{T\}_\mathcal{U}$
      \STATE $\{\mathbf{h}\}_\mathcal{U} \leftarrow$ get last element from $H_i$
      \IF{$\mathsf{AC.Evaluate}(\{\mathbf{t}\}_\mathcal{U}=\{\mathbf{h}\}_\mathcal{U})$}
      \FOR{each vertex $\{\mathbf{v}\}_\mathcal{U} \in \{T\}_\mathcal{U}$}
      \STATE (\textit{party $u$ broadcasts:})   $\mathrm{thisUnit}.\mathbf{v}\_post\_d \leftarrow \mathrm{thisUnit}.\mathbf{v}\_post\_d + 1$ 
      \ENDFOR
       \STATE (\textit{party $v$ broadcasts:}) $(\mathrm{Unit})$ \\
 $\mathbf{q}\_seq, \mathbf{q}\_post\_d, \mathbf{q}\_par\_b, \mathbf{q}\_e \leftarrow |H_i|, 0,\O, 0$
        \STATE $H_i \leftarrow  H_i\bigcup \mathrm{thisUnit}.\{\mathbf{q}\}_\mathcal{U}$   
             \IF{$\{W_i\}_\mathcal{U}/\{T\}_\mathcal{U} \neq \O$}
          \STATE $S \leftarrow S\bigcup \{W_i\}_\mathcal{U}/\{T\}_\mathcal{U} $ 
        \ENDIF
      \ENDIF
      \IF{ $\{\mathbf{t}\}_\mathcal{U} \neq \{\mathbf{h}\}_\mathcal{U}$ }
          \STATE $S \leftarrow S\bigcup \{W_i\}_\mathcal{U}$  
        \ENDIF
        \FOR{each vertex $\{\mathbf{v}\}_\mathcal{U} \in S$}
       \STATE $H_j \leftarrow$ instantiate a new branch  
     \STATE $\mathrm{thisEntry}.\mathbf{v}\_par\_b \leftarrow \mathrm{thisEntry}.\mathbf{v}\_par\_b \bigcup loc\_H_j$ // record parallel branches location of $\{\mathbf{v}\}_\mathcal{U}$ in $H_i$
      \STATE $(\mathrm{Entry})\ \mathbf{v}\_seq, \mathbf{v}\_post\_d, \mathbf{v}\_par\_b \leftarrow 0,1,\O$
     \STATE $H_j \leftarrow H_j \bigcup \mathrm{thisEntry}.\{\mathbf{v}\}_\mathcal{U}$
        \STATE $(\mathrm{Entry}) \ \mathbf{q}\_seq, \mathbf{q}\_post\_d, \mathbf{q}\_par\_b \leftarrow 1,0,\O$
      \STATE $H_j \leftarrow H_j \bigcup \mathrm{thisEntry}.\{\mathbf{q}\}_\mathcal{U}$
     \ENDFOR      
    \end{algorithmic}
\end{algorithm}

\begin{algorithm}[h]
\renewcommand{\thealgocf}{\ref{apx:sp_a2nn_algorithms}.3}
\caption{$\mathsf{Seach\text{-}Layer.HoneycombNeighbors}(\{\mathbf{c}\}_\mathcal{U}, H_i)$} 
\label{alg:search_layer.honeycomb_neighbors}
 \begin{algorithmic}[1]
\renewcommand{\algorithmicrequire}{\textbf{Input}:} 
      \REQUIRE a vertex  $\{\mathbf{c}\}_\mathcal{U}$,  its located branch $H_i$
\renewcommand{\algorithmicrequire}{\textbf{Output}:} 
      \REQUIRE neighbors of vertex $\{\mathbf{c}\}_\mathcal{U}$
\renewcommand{\algorithmicrequire}{\textbf{Finding Neighbors across Branches} ---} 
      \REQUIRE
      \STATE $Neighbors \leftarrow \O$ // set of neighbors of vertex $\{\mathbf{c}\}_\mathcal{U}$
    \STATE $\mathbf{c}\_par\_b \leftarrow $ get the parallel 
    branches set of $\mathbf{c}$ in $H_i$
    \STATE // if $\mathbf{c}$ is head vertex in $H_i$,\\ \quad \ $seq$ in line 5 starts from $\mathbf{c}.seq+1$ to $\mathbf{c}.post\_d$
     \STATE // if $\mathbf{c}$ is tail vertex in $H_i$,\\ \quad \ $seq$ in line 5 is assigned $\mathbf{c}.seq-1$
    \FOR{$seq\leftarrow  \mathbf{c}.seq-1, \mathbf{c}.seq+1 \ ...\ \mathbf{c}.seq + \mathbf{c}.post\_d $ in $H_i$} 
    \STATE $(\{\mathbf{v}\}_\mathcal{U}, loc\_H_i) \leftarrow$ get the $seq^{th}$ vertex in $H_i$
    \STATE $Neighbors =  Neighbors \bigcup (\{\mathbf{v}\}_\mathcal{U}, loc\_H_i)$
    \ENDFOR
    
    \FOR{each branch $H_j$ in $\mathbf{c}\_par\_b$}
   \STATE // get the next vertex of head vertex in $H_j$
    \STATE $(\{\mathbf{v}\}_\mathcal{U}, loc\_H_j) \leftarrow$ get the $1^{st}$ vertex of $H_j$
    \STATE $Neighbors =  Neighbors \bigcup (\{\mathbf{v}\}_\mathcal{U}, loc\_H_j)$
    \ENDFOR
    \RETURN $Neighbors$
\end{algorithmic} 
\end{algorithm}

\begin{algorithm}[h]
\renewcommand{\thealgocf}{\ref{apx:sp_a2nn_algorithms}.5}
\caption{$\mathsf{SP\text{-}A^2NN.Search}(K_\mathsf{SS,AC}, \sigma, \{\mathbf{q}\}_{\mathcal{U}}, \theta; \\
C\text{-}EDB)$} 
\label{alg:sp_a2nn.search}
 \begin{algorithmic}[1]
\renewcommand{\algorithmicrequire}{\textbf{Input}:} 
      \REQUIRE a query $\{\mathbf{q}\}_{\mathcal{U}}$ submitted by party $v$,
      maximum nearest neighbor number $\theta$; \\   $C\text{-}EDB$:  multiple bitgraphs, and its an enter vector $\{\mathbf{ev}\}_{\mathcal{U}}$ shared from party $u$, which is  located in branch $H_a$ of top layer's bitgraph (i.e., the $L^{th}$ layer). \\
      \textit{($\theta$ and locations of bitgraph, branches, and units are public parameters.)}
\renewcommand{\algorithmicrequire}{\textbf{Output}:} 
      \REQUIRE  $\theta$ nearest neighbor vectors to $\{\mathbf{q}\}_{\mathcal{U}}$. 
\renewcommand{\algorithmicrequire}{\textbf{Search Procedure} ---} 
      \REQUIRE
      \STATE $(\{\mathbf{ev}\}_{\mathcal{U}}, loc\_H_a) \leftarrow$ get enter vector for $C\text{-}EDB$
      \FOR{$l \leftarrow L ... 1$}
      \STATE $\{W\}_\mathcal{U} \leftarrow \mathsf{Search\text{-}Layer}($ \\ \quad \quad \quad \quad $\{\mathbf{q}\}_{\mathcal{U}}, \theta=1;  C\text{-}E\mathcal{I}\text{-}l, (\{\mathbf{ev}\}_{\mathcal{U}}, loc\_H_a))$
      \STATE $(\{\mathbf{ev}\}_{\mathcal{U}}, loc\_H_a) \leftarrow$ get first element from $\{W\}_\mathcal{U}$ 
      \ENDFOR
      \STATE $\{W\}_\mathcal{U} \leftarrow \mathsf{Search\text{-}Layer}($ \\  \quad \quad \quad   $\{\mathbf{q}\}_{\mathcal{U}}, \theta;  C\text{-}ED, (\{\mathbf{ev}\}_{\mathcal{U}}, loc\_H_a))$ // $0^{th}$ layer
      \STATE  $W \leftarrow \mathsf{AC.Evaluate(SS.Recon}(\{W\}_\mathcal{U}))$
      \RETURN $W$
\end{algorithmic} 
\end{algorithm}

\begin{algorithm}[h]
\renewcommand{\thealgocf}{\ref{apx:sp_a2nn_algorithms}.4}
\caption{$\mathsf{Search\text{-}Layer}(\{\mathbf{q}\}_{\mathcal{U}}, \theta; $\\ $
C\text{-}E\mathcal{I}\text{-}l, (\{\mathbf{ev}\}_{\mathcal{U}}, loc\_H_a))$}
\label{alg:search_layer}
    \begin{algorithmic}[1]
\renewcommand{\algorithmicrequire}{\textbf{Clients Input}:} 
      \REQUIRE a query $\{\mathbf{q}\}_{\mathcal{U}}$ submitted by party $v$, maximum nearest neighbor number $\theta$; \\$C\text{-}E\mathcal{I}\text{-}l, (\{\mathbf{ev}\}_{\mathcal{U}}, loc\_H_a)$:  a bitgraph of layer $l$, and an enter vector $\{\mathbf{ev}\}_{\mathcal{U}}$ with its location $loc\_H_a$. 
\renewcommand{\algorithmicrequire}{\textbf{Clients Output}:} 
      \REQUIRE nearest neighbor vectors to $\{\mathbf{q}\}_{\mathcal{U}}$.
\renewcommand{\algorithmicrequire}{\textbf{Search Procedure on a Layer}:} 
      \REQUIRE
      \renewcommand{\algorithmicrequire}{\textit{party $u$}:} 
      \REQUIRE
     \STATE $\{\mathbf{C}\}_\mathcal{U} \leftarrow (\{\mathbf{ev}\}_\mathcal{U}, loc\_H_a)$ // queue of candidates and their branch locations
      \STATE $\{\mathbf{W}\}_\mathcal{U} \leftarrow (\{\mathbf{ev}\}_\mathcal{U}, loc\_H_a)$ // queue of found nearest neighbors
  \renewcommand{\algorithmicrequire}{\textit{all parties in $\mathcal{U}$}:} 
      \REQUIRE
      \WHILE{$|C|>0$}
       \STATE $(\{\mathbf{c}\}_\mathcal{U}, loc\_H_i) \leftarrow $ extract first element from  $\{\mathbf{C}\}_\mathcal{U}$
       \STATE $\{\mathbf{f}\}_\mathcal{U} \leftarrow  $   get last element from $\{\mathbf{W}\}_\mathcal{U}$
       \IF{$\mathsf{AC.Distance}(\{\mathbf{c}\}_\mathcal{U}, \{\mathbf{q}\}_\mathcal{U}) >  \mathsf{AC.Distance}(\{\mathbf{f
       }\}_\mathcal{U}, \{\mathbf{q}\}_\mathcal{U})$}
    \STATE \textbf{break}
      \ENDIF
      \WHILE{  $\mathsf{AC.Evaluate}(\mathsf{SS.Recon}(\{\mathbf{c}\_post\_d\}_\mathcal{U})=0)$}
      \STATE remove first element from  $\{\mathbf{C}\}_\mathcal{U}$
      \STATE  $(\{\mathbf{c}\}_\mathcal{U}, loc\_H_i) \leftarrow $ extract first element from  $\{\mathbf{C}\}_\mathcal{U}$
      \ENDWHILE
    \FOR{ each $(\{\mathbf{v}\}_\mathcal{U}, loc\_H_j) \in \mathsf{Search\text{-}Layer.HoneycombNeighbors}(\{\mathbf{c}\}_\mathcal{U}, H_i)$ }
    \STATE (\textit{party $u$ broadcasts:}) $\mathrm{thisUnit}.\mathbf{v}\_e \leftarrow 1$ // record this vertex as `\textit{evaluated}'
    \STATE $\{\mathbf{f}\}_\mathcal{U} \leftarrow$ get last element from $\{W\}_\mathcal{U}$
      \IF{$\mathsf{AC.Distance}(\{\mathbf{v}\}_\mathcal{U}, \{\mathbf{q}\}_\mathcal{U}) <  \mathsf{AC.Distance}(\{\mathbf{f
       }\}_\mathcal{U}, \{\mathbf{q}\}_\mathcal{U})$}
    \STATE $\{\mathbf{C}\}_\mathcal{U} \leftarrow \{\mathbf{C}\}_\mathcal{U} \bigcup (\{\mathbf{v}\}_\mathcal{U}, loc\_H_j)$
    \STATE $\{\mathbf{W}\}_\mathcal{U} \leftarrow \{\mathbf{W}\}_\mathcal{U} \bigcup (\{\mathbf{v}\}_\mathcal{U}, loc\_H_j)$
    \IF{$\mathsf{AC.Agree}(|\{\mathbf{W}\}_\mathcal{U}|>\theta)$}
    \STATE remove last element of $\{\mathbf{W}\}_\mathcal{U}$
    \ENDIF
      \ENDIF    
    \ENDFOR    
       \ENDWHILE
    \end{algorithmic}
\end{algorithm}

\end{appendices}

%\ifCLASSOPTIONcompsoc \section*{Acknowledgments} \else \section*{Acknowledgment} \fi

\end{document}